\documentclass[12pt]{article}
\usepackage{amsmath,empheq}
\usepackage{geometry}
\usepackage{graphicx}
\usepackage{placeins}
\usepackage{seqsplit}
\usepackage{physics}
\usepackage{mhchem}
\usepackage{subcaption}
\usepackage{mathptmx}
\usepackage{setspace}
\usepackage{enumitem}
\usepackage{comment}
\usepackage{titlesec}
\usepackage{float}
\usepackage{todonotes}
\usepackage{amssymb}
\usepackage{tikz}
\usetikzlibrary{decorations.pathreplacing}
\usepackage[dotinlabels]{titletoc}
\numberwithin{equation}{section}
\titleformat{\section}[hang]{\bfseries\large}{\thesection.}{0.4em}{}
\titlelabel{\thetitle.\quad}
\graphicspath{{.}{figures/}}
\title{\textbf{Multi-Output Physics-Informed Neural Networks for Forward and Inverse PDE Problems with Uncertainties}}
\author{Mingyuan Yang$^*$ \and John T. Foster$^*$}
\date{$^*$ Department of Petroleum and Geosystems Engineering, The University of Texas at Austin}
\begin{document}
\maketitle
\singlespacing

\section*{Abstract}

Physics-informed neural networks (PINNs) have recently been used to solve various computational problems which are governed by partial differential equations (PDEs).  In this paper, we propose a multi-output physics-informed neural network (MO-PINN) which can provide solutions with uncertainty distributions for both forward and inverse PDE problems with noisy data.  In this framework, the uncertainty arising from the noisy data is first translated into multiple measurements regarding the prior noise distribution using the bootstrap method, and then the outputs of neural networks are designed to satisfy the measurements as well as the underlying physical laws.  The posterior estimation of target parameters can be obtained at the end of training, which can be further used for uncertainty quantification and decision making.  In this paper, MO-PINNs are demonstrated with a series of numerical experiments including both linear and nonlinear, forward and inverse problems.  The results show that MO-PINN is able to provide accurate predictions with noisy data.

In addition,  we also demonstrate that the prediction and posterior distributions from MO-PINNs are consistent with the solutions from traditional a finite element method (FEM) solver and Monte Carlo methods given the same data and prior knowledge.  Finally, we show that additional statistical knowledge can be incorporated into the training to improve the prediction if available.

\section{Introduction}
Deep learning techniques have been developed over the past few years to solve a range of computational science and engineering problems \cite{berg2018unified,goswami2019transfer,haghighat2020deep,he2020physics,tchelepi2020limitations}.  Unlike traditional applications of deep learning where a large datasets are needed to train the models \cite{raissi2018deep,raissi2019physics}, physics-informed neural networks (PINNs) have been developed to embed physical balance laws into the loss function so that a smaller number of data/measurements are required to achieve an accurate model \cite{raissi2019physics,zhu2019physics}. 

Futhermore, researchers have been trying to extend the capabilities of PINNs so that they can be used in scenarios with noisy measurements for uncertainty quantification.  In \cite{li2014adaptive,yan2019adaptive}, Bayesian inference was used to quantify uncertainties in inverse PDE problems.  In \cite{yang2018physics}, generative adversarial networks (GANs) were used to quantify uncertainties in stochastic differential equations (SDEs).  Bayesian physics-informed neural networks (BPINNs) were proposed in \cite{yang2021b,meng2021multi} to solve PDEs for both forward and inverse problems with noisy data.  Our work is inspired by results with BPINNs as the problem setting with sparse noisy measurements is quite common in engineering practice. 

In this work, we modify the structure of PINNs which was proposed in \cite{raissi2019physics} and show that the outputs from the neural network are capable of reflecting the distributions of noisy data while making a reasonable predictions of unknown parameters.  We demonstrate our approach on the same forward and inverse problems as in \cite{yang2021b}, so that the readers may compare the results between the two approaches. The remainder of this paper is organized as follows: in Section~\ref{MOU-PINNs}, we briefly introduced the MO-PINNs and its loss function with respect to noisy data, in Section~\ref{results}, we present the numerical results for forward problems, inverse problems, and a case with additional physical constraints, in Section~\ref{summary}, we briefly talk about the limitation of our approach and the possible applications beyond the current work.

\section{Multi-Output Physics-Informed Neural Networks} \label{MOU-PINNs}
In the framework of PINNs, the conservation equations, constitutive laws, and other physical constraints are embedded into the overall loss function.  The loss function is minimized by using optimizers in an iterative fashion so that the model parameters are updated.  At the end of training, a neural network which satisfies the prescribed constraints is obtained and it can return a value of the target function at any given point in the domain of interest.

MO-PINNs are a direct extension of PINNs, but use multiple neurons in the output layer so that any knowledge regarding uncertainty can be imposed on the distribution which is formed by outputs at each point.

\begin{figure}[H]
    \begin{center}
        \begin{tikzpicture}[shorten >=1pt]
            \tikzstyle{unit}=[draw,shape=circle,minimum size=1.15cm]
            \node[unit, fill={rgb:orange,1;yellow,2;pink,5}](x) at (0,3.5){$x$};
            \node[unit, fill={rgb:orange,1;yellow,2;pink,5}](y) at (0,2){$y$};
            \node[unit, fill={rgb:orange,1;yellow,2;pink,5}](z) at (0,0.5){$z$};
            \node[unit, fill={rgb:orange,1;yellow,2;pink,5}](t) at (0,-1){$t$};
            \node[unit, fill={rgb:orange,1;yellow,2;pink,5}](y1) at (3,2.5){$y^1$};
            \node(dots) at (3,1.65){\vdots};
            \node[unit, fill={rgb:orange,1;yellow,2;pink,5}](yc) at (3,0.5){$y^N$};
    
            \node[unit, fill={rgb:red,1;green,2;blue,5}](u1) at (5,2.5){$u_{NN}^1$};
            \node(dots) at (5,1.65){\vdots};
            \node[unit, fill={rgb:red,1;green,2;blue,5}](u2) at (5,0.5){$u_{NN}^M$};
    
            \node[unit, fill={rgb:orange,1;yellow,2;pink,5}](dt) at (8,4.0){$\frac{\partial u_{NN}}{\partial t}$};
            \node[unit, fill={rgb:orange,1;yellow,2;pink,5}](grad) at (8,2.0){$\nabla \cdot u_{NN}$};
            \node[unit, fill={rgb:orange,1;yellow,2;pink,5}](lap) at (8,0.2){$\Delta u_{NN}$};
            \node[unit, fill={rgb:red,1;green,2;blue,5}](stats) at (8,-1.5){$P(u_{NN})$};
    
            \node[unit, fill={rgb:orange,1;yellow,2;pink,5}](l) at (11,1.5){$Loss$};
     
            \draw[->] (x) -- (y1);
            \draw[->] (y) -- (y1);
            \draw[->] (z) -- (y1);
            \draw[->] (t) -- (y1);
     
            \draw[->] (x) -- (yc);
            \draw[->] (y) -- (yc);
            \draw[->] (z) -- (yc);
            \draw[->] (t) -- (yc);
     
            \draw[->] (y1) -- (u1);
            \draw[->] (yc) -- (u1);
            \draw[->] (y1) -- (u2);
            \draw[->] (yc) -- (u2);
    
            \draw[->] (u1) -- (dt);
            \draw[->] (u1) -- (grad);
            \draw[->] (u1) -- (lap);
            \draw[->] (u1) -- (stats);
            \draw[->] (u2) -- (dt);
            \draw[->] (u2) -- (grad);
            \draw[->] (u2) -- (lap);
            \draw[->] (u2) -- (stats);
    
            \draw[->] (dt) -- (l);
            \draw[->] (grad) -- (l);
            \draw[->] (lap) -- (l);
            \draw[->] (stats) -- (l);
     
            \draw [decorate,decoration={brace,amplitude=10pt},xshift=-4pt,yshift=0pt] (-0.5,4) -- (0.75,4) node [black,midway,yshift=+0.6cm]{\small Input layer};
            \draw [decorate,decoration={brace,amplitude=10pt},xshift=-4pt,yshift=0pt] (2.5,3) -- (3.75,3) node [black,midway,yshift=+0.6cm]{\small Hidden layer};
            \draw [decorate,decoration={brace,amplitude=10pt},xshift=-4pt,yshift=0pt] (5.75,-0.1) -- (4.5,-0.1) node [black,midway,yshift=-0.6cm]{\small Output layer};
            \draw [decorate,decoration={brace,amplitude=10pt},xshift=-4pt,yshift=0pt] (5.95,-1.5) -- (-0.5,-1.5) node [black,midway,yshift=-0.6cm]{\small Neural networks};
            \draw [decorate,decoration={brace,amplitude=10pt},xshift=-4pt,yshift=0pt] (4.5,4.6) -- (11.5,4.6) node [black,midway,yshift=+0.6cm]{\small Loss function};
        \end{tikzpicture}
        \caption{A fully connected neural network schematic with 4, N, and M neurons in the input, hidden, and output layers respectively. The loss function is constructed with operations applied on the outputs of the neural networks $u_{NN}$.  The yellow parts of the diagram are identical in PINNs, while the blue parts indicate the multiple outputs with respect to the same inputs as well as the corresponding statistical properties $P(u_{NN})$.}
        \label{fig:nn schematic}
    \end{center}
\end{figure}
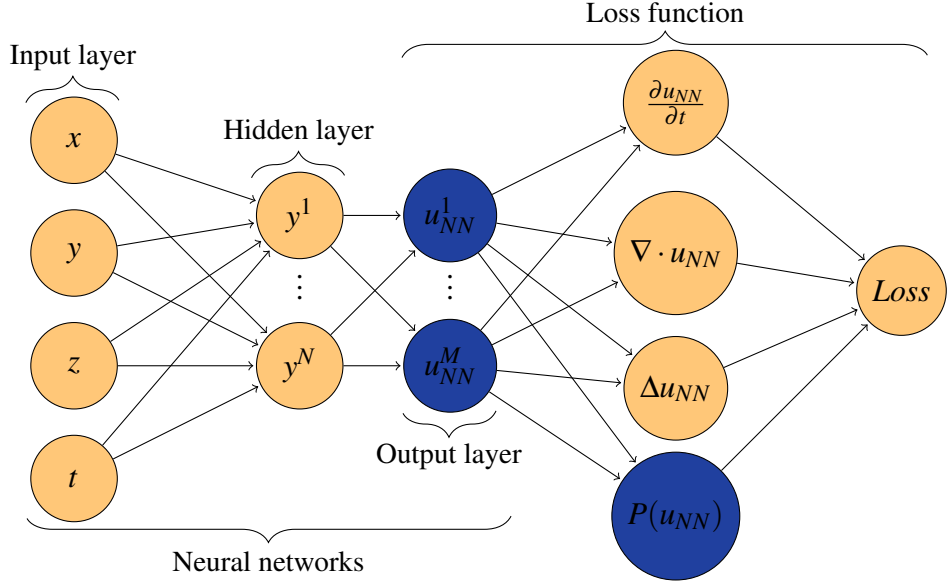

A schematic illustration of a MO-PINN is shown in Figure~\ref{fig:nn schematic} where it shows differences in construction compared to the PINNs in \cite{raissi2019physics}.  There are multiple neurons in the output layer corresponding to the same inputs $x$, $y$, $z$ and $t$ of the network.  Therefore, we can use these outputs at the same point to calculate statistical properties ($P(u_{NN})$) and impose the prior knowledge or assumptions regarding uncertainty in the loss function.

Here we consider a general inverse problem govern by PDEs and constraints for demonstration,

\begin{gather}
    \mathcal{L} u \left( x \right) = f(x),\qquad x \in \Omega, \label{general balance equation}
    \\
    u(x) = h(x),\qquad x \in \partial \Omega_h, \label{essential}
    \\
    \frac{\partial u(x)}{\partial x} = g(x),\qquad x \in \partial \Omega_g, \label{natural}
    \\
    u(x^u_i) = u_m(x^u_i), \qquad i=1,2,...,n \label{u measure}
    \\
    f(x^f_i) = f_m(x^f_i), \qquad i=1,2,...,m \label{f measure}
\end{gather}

where $\Omega$ is the domain bounded by the boundaries $\partial \Omega$, $u(x)$ is the solution to the PDEs on $\Omega$, $\mathcal{L}$ is a differential operator that is applied on $u(x)$.  $\partial\Omega_h$ is part of the boundary where essential boundary condition $h(x)$ is applied, and $\Omega_g$ is the part of boundary where the natural boundary condition $g(x)$ is applied.  $u_m$ and $f_m$ are the $n$ and $m$ noisy measurements available for training while $x^u$ and $x^f$ are respective spatial locations of the noisy data.  The objective of this inverse problem is to obtain predictions of distributions of $u$ and $f$ with respect to the PDE, constraints, and data.

Two neural networks $u_{NN}$ and $f_{NN}$ following the structure in Figure~\ref{fig:nn schematic} are constructed to represent the solution $u$ and $f$.
In this work, and without loss of generalization, only fully connected neural networks are considered which can be expressed as
\begin{gather}
    u_{NN}(x;W,b) = T^{m} \circ T^{m-1} \circ \cdot \cdot \cdot \circ T^{2} \circ T^{1}(x), \notag
    \\
    T^j(\cdot) = \sigma^j(W^j\times \cdot + b^j),\qquad  j=1,2,\ldots,n, \notag
\end{gather}
where $W^j$ and $b^j$ are weights and bias of layer $j$, $(\cdot)$ is the output from the previous layer. 
Note that the structure of neural networks yield hyperparameters that can affect the performance of training. In paper \cite{wang2020understanding}, the authors show that incorporating the attention mechanism into fully connected networks can accelerate the training.  This topic is out the scope of this work so that we will not further discuss it here.

The error in measurements are randomly drawn from respective error distributions $\sigma_u$ and $\sigma_f$, and then assigned to the outputs of neural networks $u_{NN}$ and $f_{NN}$ directly.
The strong form of residuals of \eqref{general balance equation}, \eqref{essential}, \eqref{natural}, \eqref{u measure} and \eqref{f measure} can be written as,

\begin{gather}
    r_d(u_{NN}^j, f_{NN}^j) = \mathcal{L} u_{NN}^j - f_{NN}^j,\qquad x \in \Omega \label{r general balance equation}
    \\
    r_e(u_{NN}^j) = u_{NN}^j - h,\qquad x \in \partial \Omega_h \label{r essential}
    \\
    r_n(u_{NN}^j) = \frac{\partial u_{NN}^j}{\partial x} - g,\qquad x \in \partial \Omega_g \label{r natural}
    \\
    r_{um}(u_{NN}^j) = u_{NN}^j(x^u_i) - \left(u_m(x^u_i) + \sigma_u^j\right), \qquad i=1,2,...,n \label{u measure residual}
    \\
    r_{fm}(f_{NN}^j) = f_{NN}^j(x^f_i) - \left(f_m(x^f_i) + \sigma_f^j\right), \qquad i=1,2,...,m \label{f measure residual}
\end{gather}

where $j=1,2,...M$ correspond to the number of neurons in the output layer of $u_{NN}$ and $f_{NN}$.
Note that there are multiple ways to approximate the residual in \eqref{r general balance equation}, and it has been extensively discussed in \cite{kharazmi2021hp}.  In this work, we use a finite difference discretization to approximate the strong form residual of \eqref{r general balance equation} because of the ease of its implementation.

The overall loss function can be written as
\begin{gather}
    L = \sum_{j=1}^{M} \left( \sum_{i}^N r_d + \sum_{i=1}^{N_e} r_e + \sum_{i=1}^{N_n} r_n + \sum_{i=1}^{n} r_{um} + \sum_{i=1}^m r_{fm}  \right),
\end{gather}
where $N$ is the number of collocation points in the domain, $N_e$ and $N_n$ are the number of collocation points on the boundaries with essential and natural boundary conditions, $n$ and $m$ are the number of noisy measurements of $u$ and $f$. The two trained networks will return $M$ values of $u$ and $f$ at each point in the domain,  so that the mean and standard deviations of the distribution of predictions across the whole domain can be further calculated.

\section{Numerical Results} \label{results}

In this section, we present several numerical experiments on both forward and inverse problems that demonstrate the utility of MO-PINNs for both linear and nonlinear equations with noisy measurements.

\subsection{Forward PDE problems}

We define \emph{forward problems} as those where all constitutive parameters are defined and we are seeking only solutions to the PDE.  Examples for linear and nonlinear problems in one and two dimensions are demonstrated.

\subsubsection{1D linear Poisson equation} \label{linear case}

Here we consider the following one-dimensional linear Poisson equation
\begin{gather}
    \lambda \frac{\partial^2 u}{\partial x^2} = f, \qquad x \in [-0.7, 0.7], \label{1d linear possion equation} 
\end{gather}
where $\lambda = 0.01$ and the solution of $u$ is assumed to be $u=\sin^3(6x)$. Therefore, the source term $f$ can be derived from the Equation~\eqref{1d linear possion equation}, i.e.\ utilizing the method of manufactured solutions. The solution $u$ and the exact source term $f$ are shown in Figure~\ref{fig:1d_linear_possion_u_and_f}.
\begin{figure}[H]
    \centering
    \subfloat[\centering Solution $u$]{{\includegraphics[width=7cm]{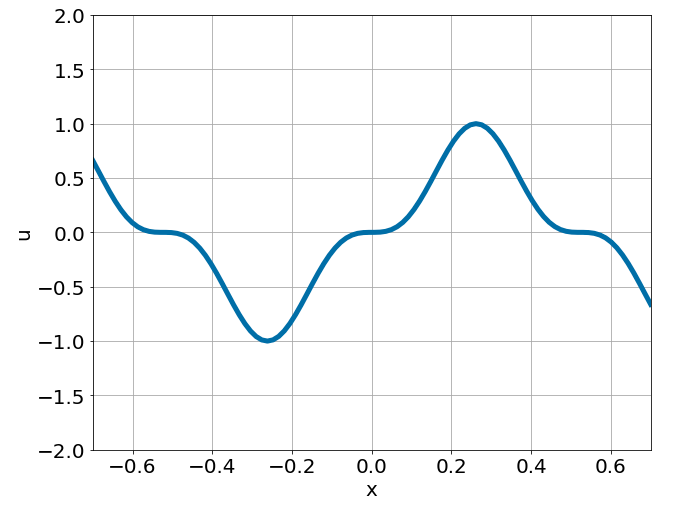} }}%
    \qquad
    \subfloat[\centering Source $f$]{{\includegraphics[width=7cm]{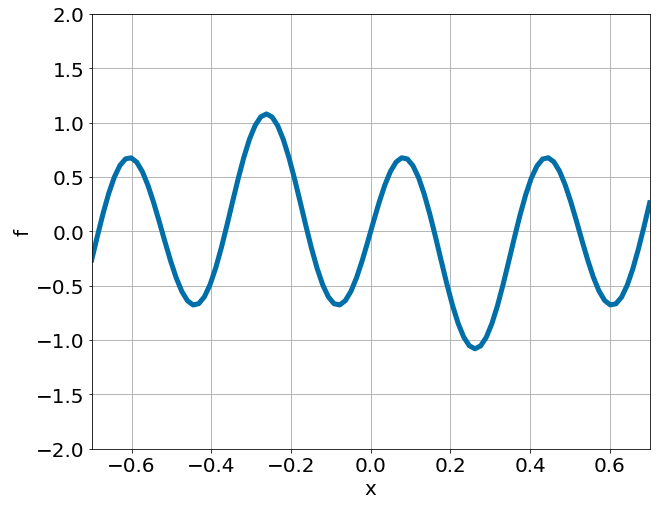} }}%
    \caption{One dimensional linear Poisson equation, solution $u$ and source $f$.}
    \label{fig:1d linear possion u and f}
\end{figure}

Now we assume that the functional form of $u$ and $f$ are both unknown, and only sparse noisy measurements of $u$ and $f$ are available as training data. The goal of training is to obtain two neural network representations for $u$ and $f$ over the whole domain.

Here we assume that 16 measurements of $f$ are available which are equally sampled in the range $x \in [-0.7, 0.70]$.  Additionally, we assume that two measurements of $u$ at two ends are available, which serve as the Dirichlet boundary conditions for the equation.  We also assume that both measurements of $u$ and $f$ are noisy and the errors follow a given distribution.  In this example, we investigate two cases where errors follow the Gaussian distribution with different standard deviations:
\begin{center}
    \begin{enumerate}
    \item $\epsilon_u \sim \mathcal{N}(0,\,0.01^{2})\,,\qquad \epsilon_f \sim \mathcal{N}(0,\,0.01^{2}),$
    \item $\epsilon_u \sim \mathcal{N}(0,\,0.1^{2})\,,\qquad \epsilon_f \sim \mathcal{N}(0,\,0.1^{2}),$
\end{enumerate}
\end{center}
The training data of $u_{data}$ and $f_{data}$ is then
\begin{gather}
    u_{data} = u_{exact} + \epsilon_u \label{u data}, 
    \\
    f_{data} = f_{exact} + \epsilon_f \label{f data}.
\end{gather}
Consequently, the multiple outputs of neural networks are
\begin{gather}
    u_{output} = u_{data} + \epsilon_u, \notag
    \\
    f_{output} = f_{data} + \epsilon_f. \notag
\end{gather}

In this work, all the examples of MO-PINNs are implemented with PyTorch \cite{paszke2017automatic}. The neural network structures are the same for $u_{NN}$ and $f_{NN}$, which are defined as fully connected neural networks with two hidden layers consisting of 20 and 40 neurons individually while the output layer consists of 500 neurons.
The hyperbolic tangent function ($\tanh{}$) is used as activation function in this section and throughout the rest of this work.
The ADAM optimizer is used for training with a learning rate of $10^{-3}$. The Xavier normal initialization strategy is used in this study which it yields neural networks with different weights at the beginning of training for different random seeds. All results in this section were obtained after 10000 epochs of training. %
\begin{figure}[H]
    \centering
    \subfloat[\centering Prediction of $u$ with raw solutions]{{\includegraphics[width=7cm]{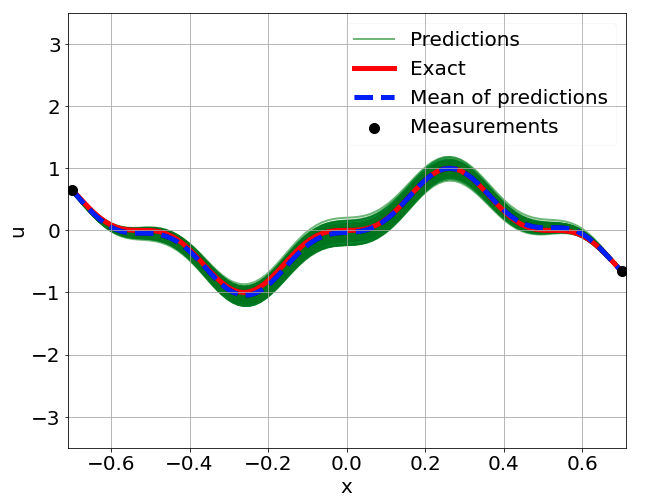} }}%
    \qquad
    \subfloat[\centering Prediction of $f$ with raw solutions]{{\includegraphics[width=7cm]{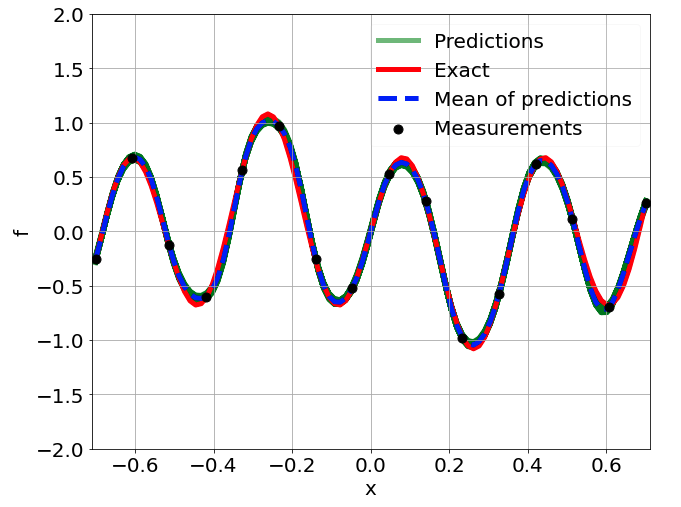} }}%
    \qquad
    \subfloat[\centering Prediction of $u$ with shady uncertain area]{{\includegraphics[width=7cm]{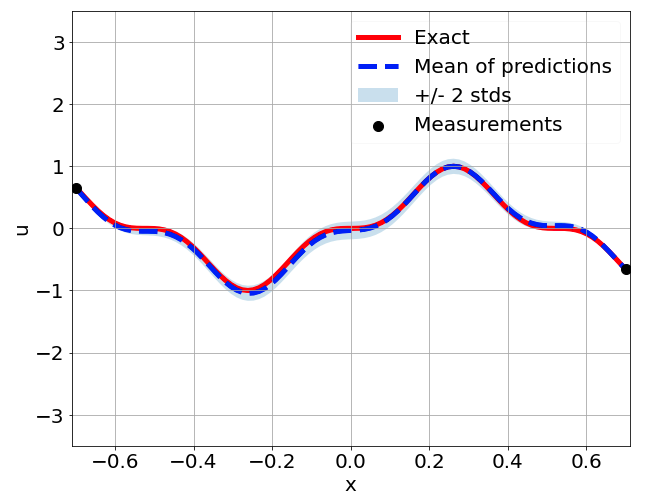} }}%
    \qquad
    \subfloat[\centering Prediction of $f$ with shady uncertain area]{{\includegraphics[width=7cm]{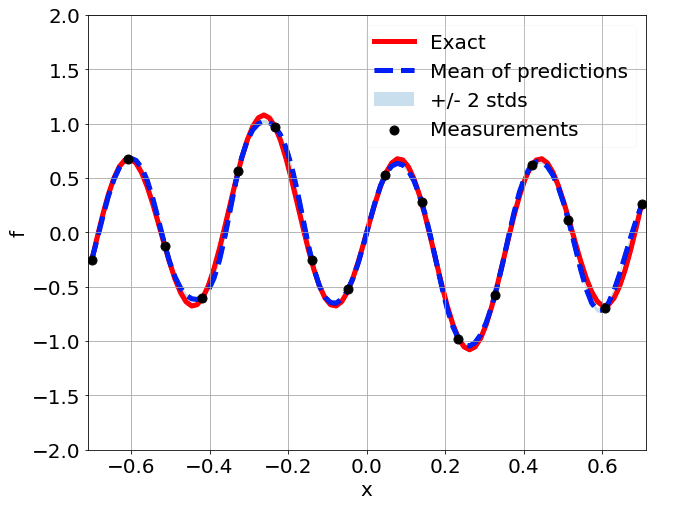} }}%
    \caption{1D linear Poisson equation, predictions of $u$ and source $f$ for case 1. The green lines are the 500 predictions, the dashed 
    blue lines are the mean of the predictions, the red lines are the exact solutions, the light blue shady areas are the mean predictions plus/minus two standard deviations and the black dots are the training data.}
    \label{fig:1d linear possion u and f prediction 0.01 noise}
\end{figure}
\begin{figure}[H]
    \centering
    \subfloat[\centering Prediction of $u$ with raw solutions]{{\includegraphics[width=7cm]{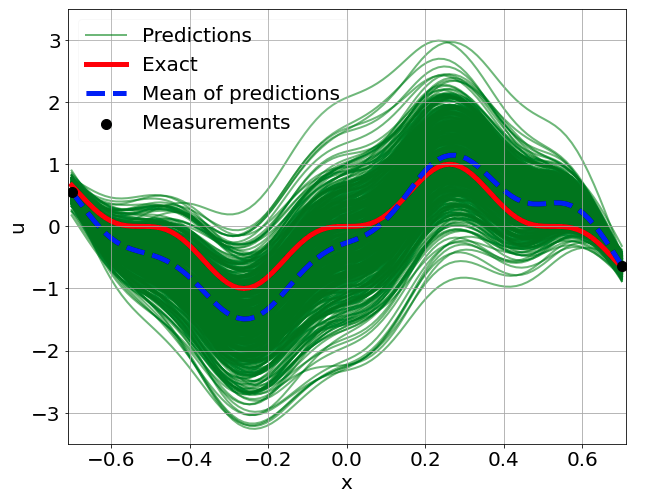} }}%
    \qquad
    \subfloat[\centering Prediction of $f$ with raw solutions]{{\includegraphics[width=7cm]{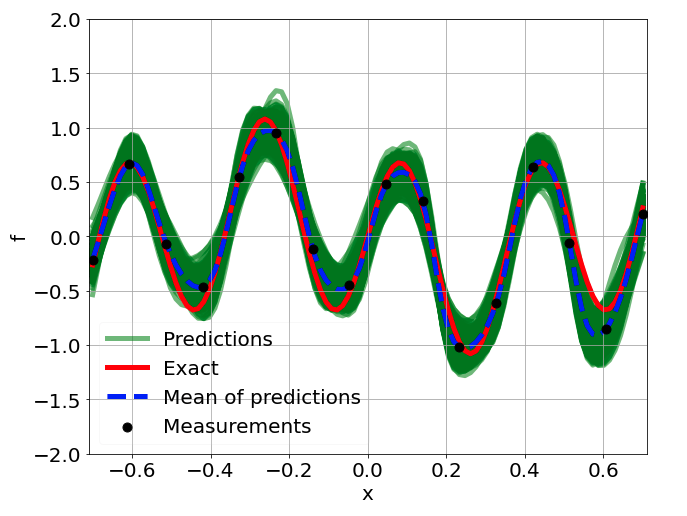} }}%
    \qquad
    \subfloat[\centering Prediction of $u$ with shady uncertain area]{{\includegraphics[width=7cm]{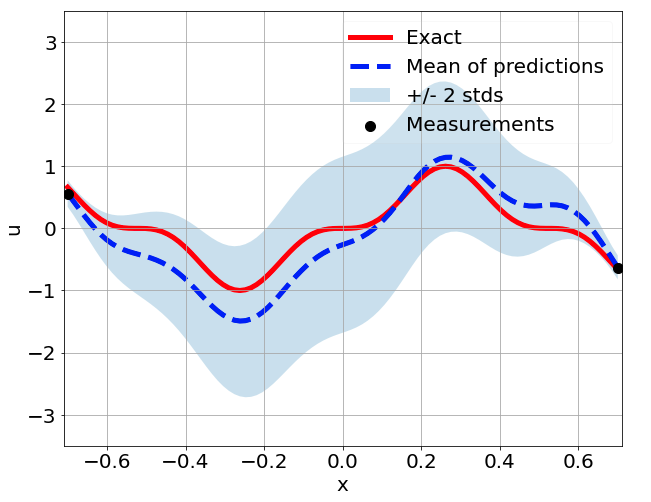} }}%
    \qquad
    \subfloat[\centering Prediction of $f$ with shady uncertain area]{{\includegraphics[width=7cm]{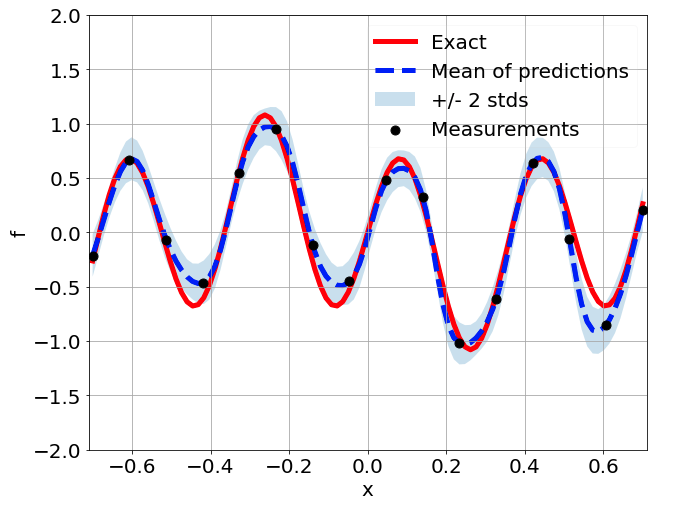} }}%
    \caption{1D linear Poisson equation, predictions of $u$ and source $f$ for case 2. The green lines are the 500 predictions, the dashed 
    blue lines are the mean of the predictions, the red lines are the exact solutions, the light blue shady areas are the mean predictions plus/minus two standard deviations and the black dots are the training data.}
    \label{fig:1d linear possion u and f prediction 0.1 noise}
\end{figure}

The results of the two cases with different noise scales are shown in Figure \ref{fig:1d linear possion u and f prediction 0.01 noise} and Figure \ref{fig:1d linear possion u and f prediction 0.1 noise}. First we see that the means of predictions for both $u$ and $f$ pass through the noisy measurements which is consistent with our error distribution assumptions.
Next we see that the multiple outputs from neural networks form the posterior estimations of $u$ and $f$ over the investigated range $[-0.7, 0.7]$.  In addition, it can be seen that the exact solutions are bounded by two standard deviations, which means that the predictions are reasonable. Last, we see that the standard deviations of predictions grow as the measurement error increases.  The results that we obtain with MO-PINN are comparable with the ones presented in paper \cite{yang2021b} using Bayesian Hamiltonian Monte Carlo physics-informed neural networks (B-PINN-HMC). However, our approach does not require assumed prior distributions of hyperparameters in neural networks and can be easily implemented by modifying the standard PINN.

\begin{figure}[H]
    \centering
    \subfloat{{\includegraphics[width=7cm]{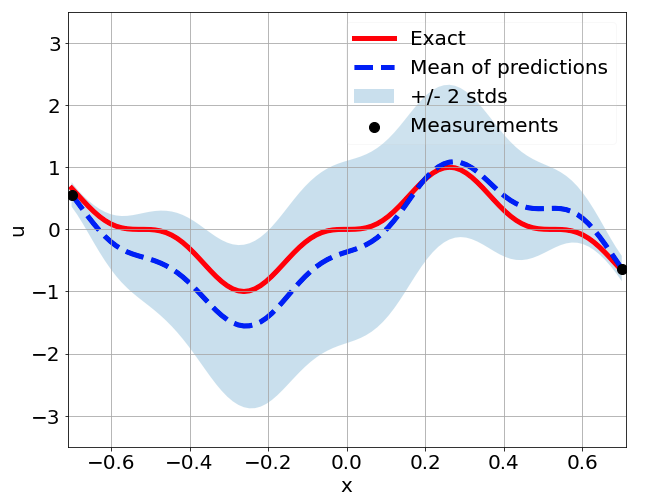} }}%
    \qquad
    \subfloat{{\includegraphics[width=7cm]{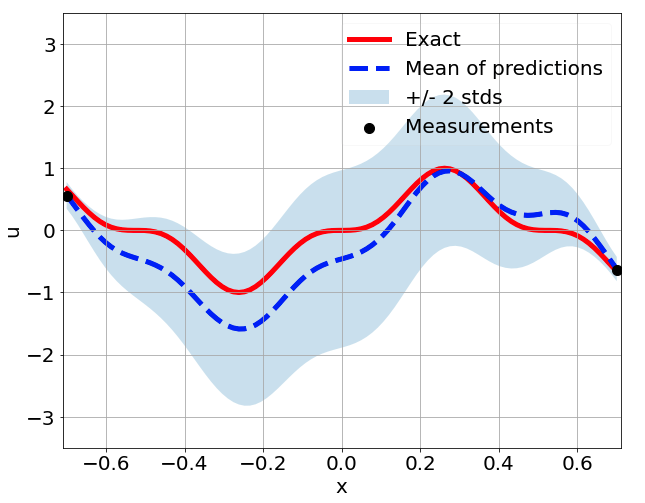} }}%
    \qquad
    \subfloat{{\includegraphics[width=7cm]{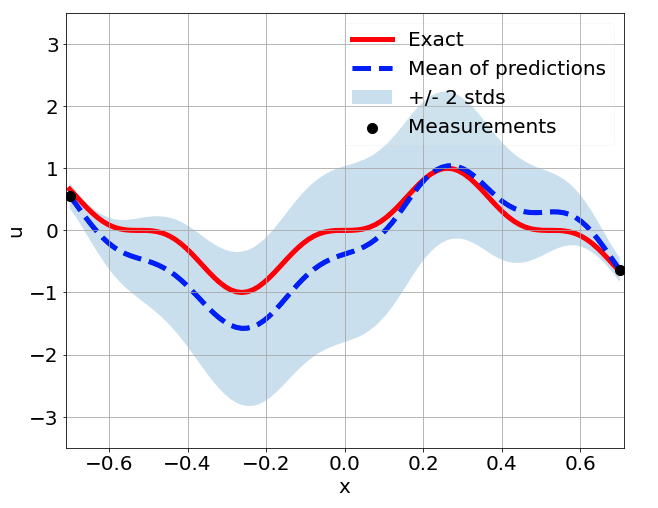} }}%
    \qquad
    \subfloat{{\includegraphics[width=7cm]{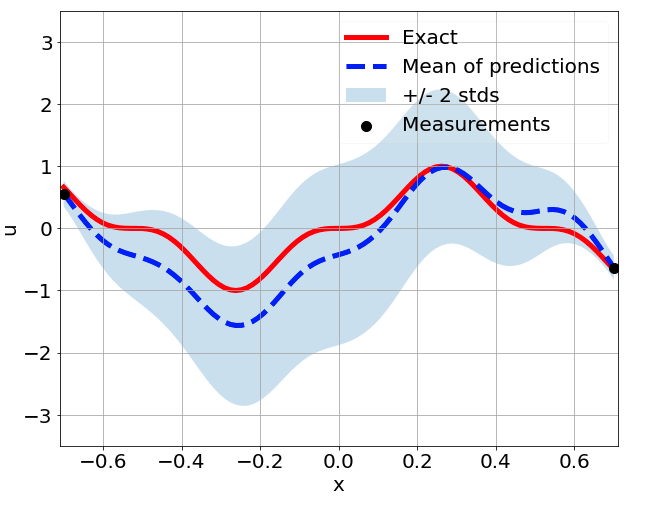} }}%
    \caption{1D linear Poisson equation, predictions of $u$ from four simulations with different random initializations and random error drawings from distributions of case 2.}
    \label{fig:1d linear possion u with different random seeds}
\end{figure}

We also verified our solutions from multiple aspects. First, we checked the consistency of the solutions when the two neural networks are initialized differently and different random noise are assigned to the outputs. Here we take the case 2 as example and run the simulations with different random seeds while keeping the measurements the same. Four predictions of $u$ are presented in Figure \ref{fig:1d linear possion u with different random seeds}. It is shown that the results are in agreement which means that the solution is consistent and invariant to the randomness at the beginning of training.

\begin{figure}[H]
    \centering
    \subfloat{{\includegraphics[width=7cm]{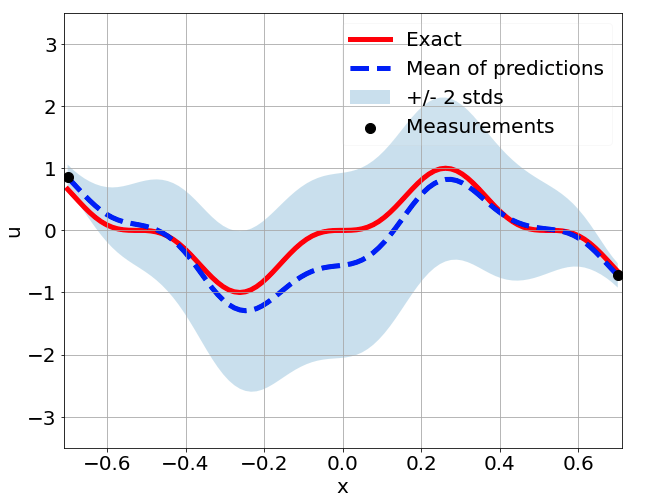} }}%
    \qquad
    \subfloat{{\includegraphics[width=7cm]{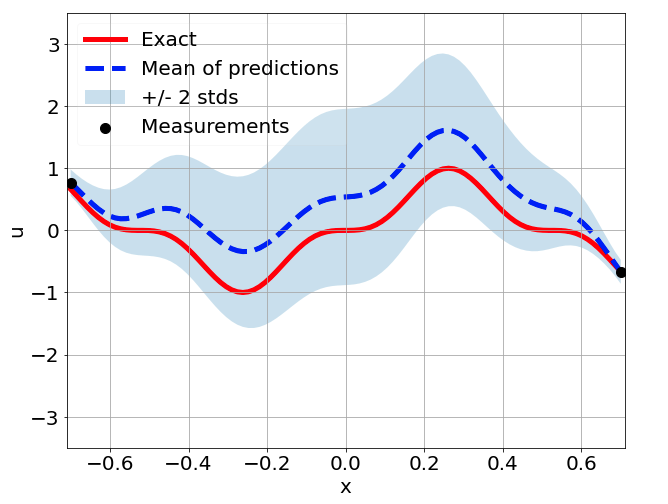} }}%
    \qquad
    \subfloat{{\includegraphics[width=7cm]{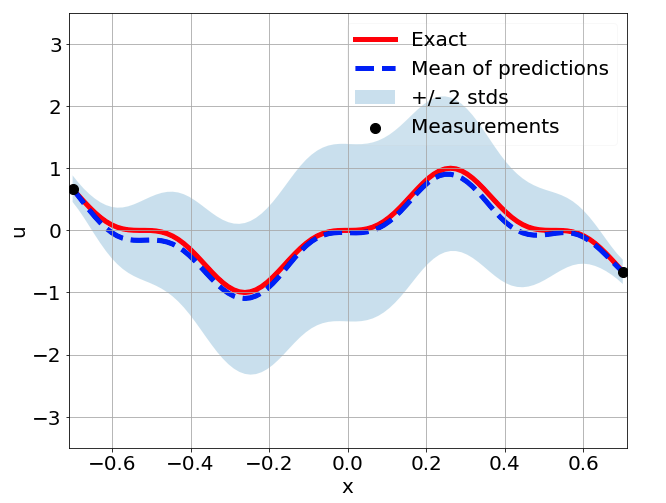} }}%
    \qquad
    \subfloat{{\includegraphics[width=7cm]{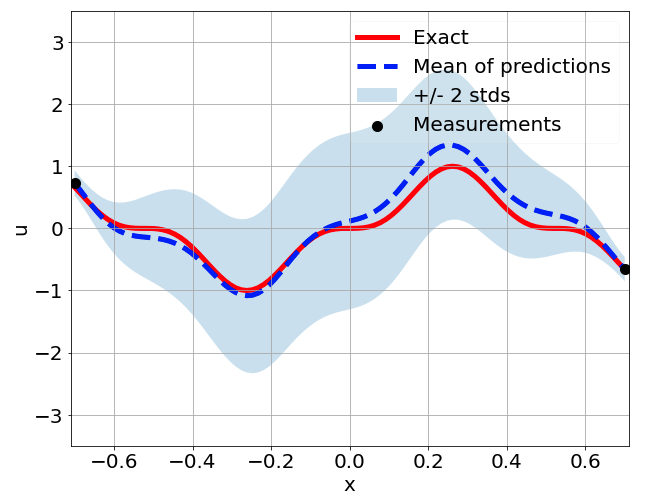} }}%
    \caption{1D linear Poisson equation, predictions of $u$ from four simulations with different measurements of case 2.}
    \label{fig:1d linear possion u with different measurements}
\end{figure}

\begin{figure}[H]
    \centering
    \subfloat[$u$ predictions with 10 outputs]{{\includegraphics[width=7cm]{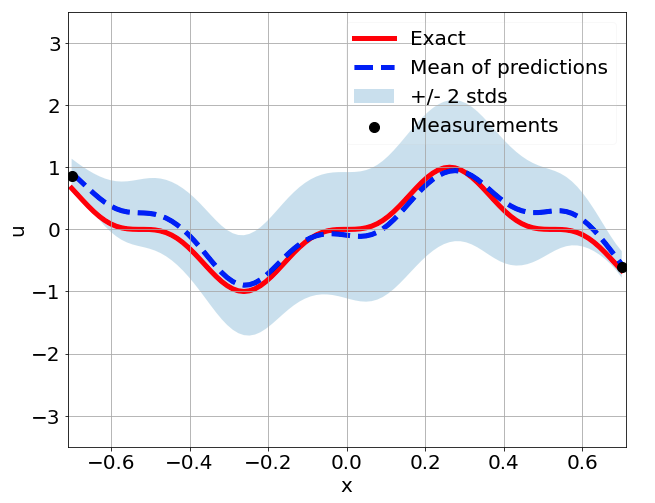} }}%
    \qquad
    \subfloat[$u$ predictions with 50 outputs]{{\includegraphics[width=7cm]{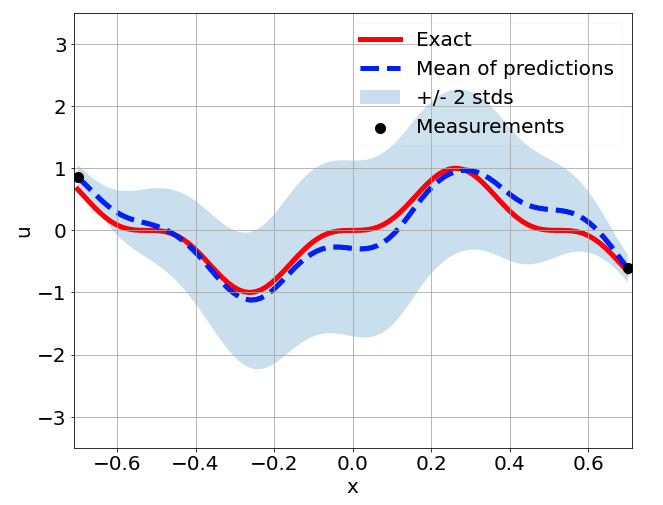} }}%
    \qquad
    \subfloat[$u$ predictions with 100 outputs]{{\includegraphics[width=7cm]{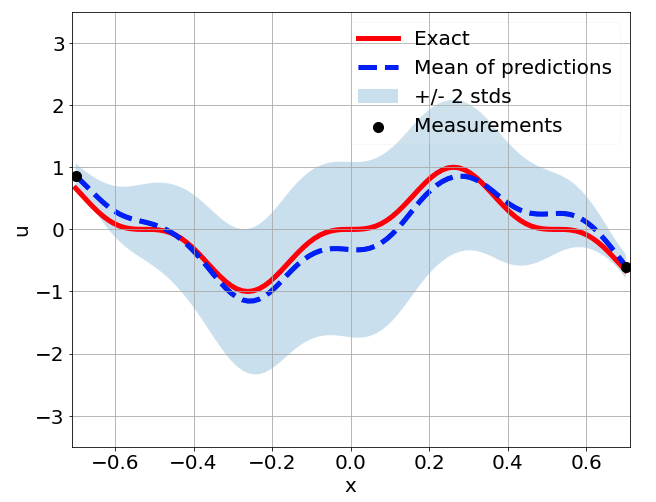} }}%
    \qquad
    \subfloat[$u$ predictions with 500 outputs]{{\includegraphics[width=7cm]{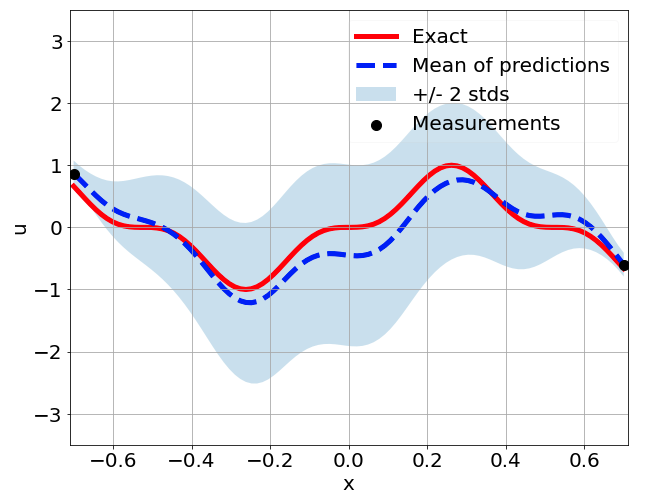} }}%
    \caption{1D linear Poisson equation, predictions of $u$ with different number of outputs for neural networks on the same measurements.}
    \label{fig:1d linear possion u with different number of outputs on the same measurements.}
\end{figure}

It should also be noted that the predictions are different with respect to different noisy measurements from Equation~\eqref{u data} and Equation~\eqref{f data}.  In an physical setting, we would likely have no control over the measurements in practice, which means that the best we can expect is that the exact solution is bounded by two or three standard deviations for all possible measurements.   Therefore, we tested our approach on different measurements and the results are shown in Figure~\ref{fig:1d linear possion u with different measurements}.  We can see that for all cases the exact solution is bounded by two standard deviations which confirms the reliability of our approach.
Finally, we also tested the convergence of the predictions with respect to different number of outputs used in the neural networks. In this work, we do not have a rigorous theory to derive the optimal number of outputs so that they can truly represent the posterior distributions. Instead, an ad hoc strategy is adopted here such that we incrementally increased the number of outputs and accepted the results when the change from previous iterations was trivial. For example, in Figure \ref{fig:1d linear possion u with different number of outputs on the same measurements.}, we tested our approach with different number of outputs for both $u$ and $f$ neural networks on the same measurements.  It is shown that the mean of predictions as well as the shady areas do not change much from 50 outputs to 100 and 500 outputs, so that we consider the predictions with 500 outputs to be our converged solution to this problem. 

\subsubsection{1D nonlinear Poisson equation} \label{1d nonlinear case}
\begin{figure}[H]
    \centering
    \subfloat[\centering Solution $u$]{{\includegraphics[width=7cm]{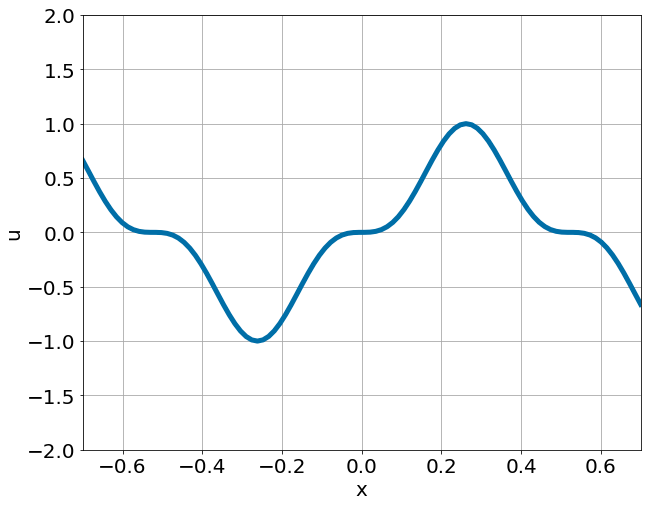} }}%
    \qquad
    \subfloat[\centering Source $f$]{{\includegraphics[width=7cm]{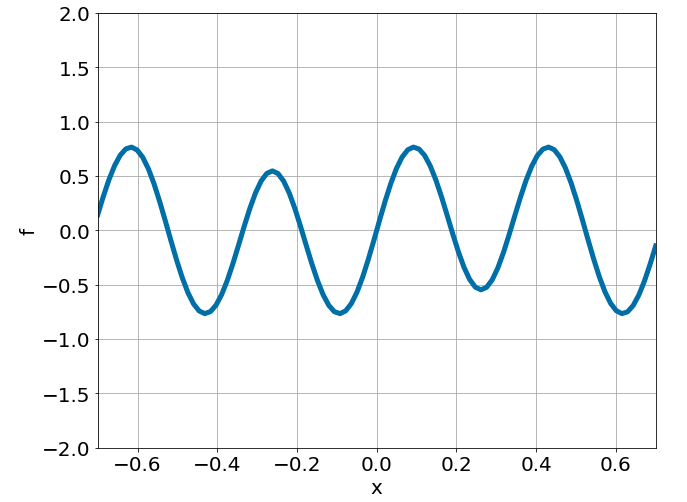} }}%
    \caption{1D nonlinear Poisson equation, solution $u$ and source $f$.}
    \label{fig:1d nonlinear possion u and f}
\end{figure}

In this section we consider the following one-dimensional nonlinear Poisson equation
\begin{gather}
    \lambda \frac{\partial^2 u}{\partial x^2} + k\tanh(u) = f, \qquad x \in [-0.7, 0.7], \label{1d nonlinear possion equation} 
\end{gather}
where $\lambda = 0.01$, $k=0.7$ and the solution $u$ is assumed to be the same with the one in the last section, i.e.\ $u=\sin^3(6x)$. Therefore, the source term $f$ can be derived from the Equation~\eqref{1d nonlinear possion equation} with manufactured solution. The solution $u$ and the exact source term $f$ are shown in Figure \ref{fig:1d nonlinear possion u and f}.

This time we assume that 32 measurements of $f$ are available to us which are equidistantly sampled in the range $x \in [-0.7, 0.7]$, while the exact expression of $f$ is assumed to be unknown. Again, we assume that two measurements of $u$ at two ends are available, which serve as the Dirichlet boundary conditions for the equation. We also investigate two cases with errors following the same distribution as in \S~\ref{linear case}. The neural network structures and the settings of training are kept the same as in last section. All results are collected after 10000 epochs. 

\begin{figure}[H]
    \centering
    \subfloat[\centering Prediction of $u$ with raw solutions]{{\includegraphics[width=7cm]{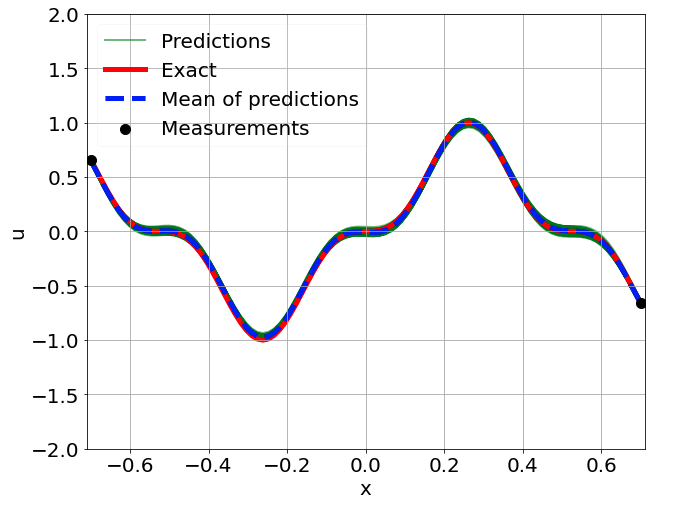} }}%
    \qquad
    \subfloat[\centering Prediction of $f$ with raw solutions]{{\includegraphics[width=7cm]{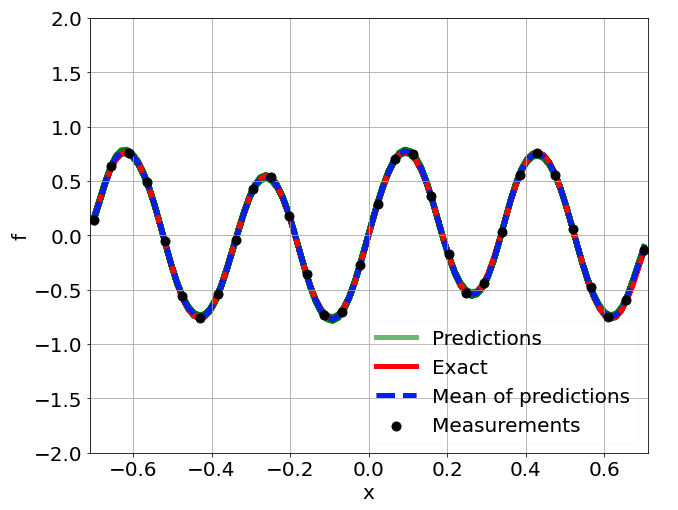} }}%
    \qquad
    \subfloat[\centering Prediction of $u$ with shady uncertain area]{{\includegraphics[width=7cm]{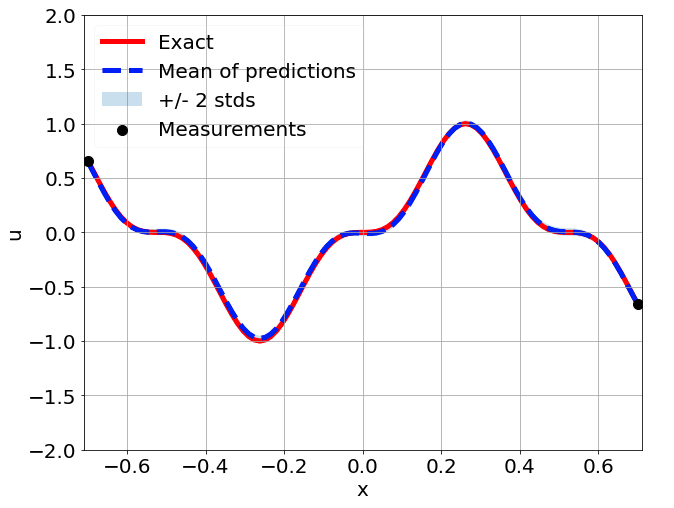} }}%
    \qquad
    \subfloat[\centering Prediction of $f$ with shady uncertain area]{{\includegraphics[width=7cm]{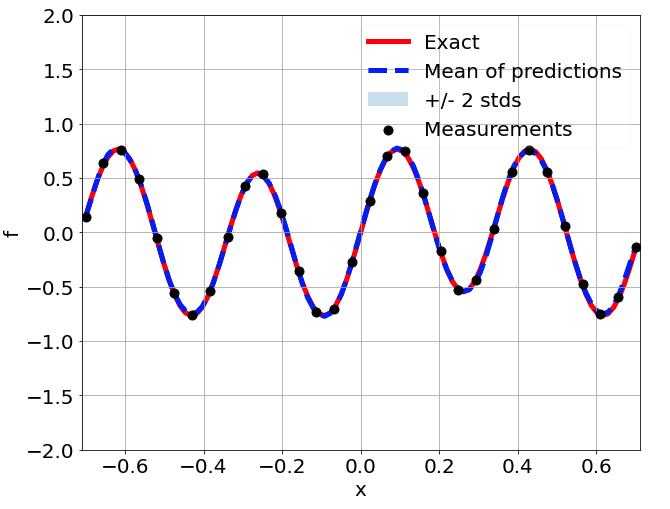} }}%
    \caption{1D nonlinear Poisson equation, predictions of $u$ and source $f$ for case 1. The green lines are the 500 predictions, the dashed 
    blue lines are the mean of the predictions, the red lines are the exact solutions, the light blue shady areas are the mean predictions plus/minus two standard deviations and the black dots are the training data.}
    \label{fig:1d nonlinear possion u and f prediction 0.01 noise}
\end{figure}

\begin{figure}[H]
    \centering
    \subfloat[\centering Prediction of $u$ with raw solutions]{{\includegraphics[width=7cm]{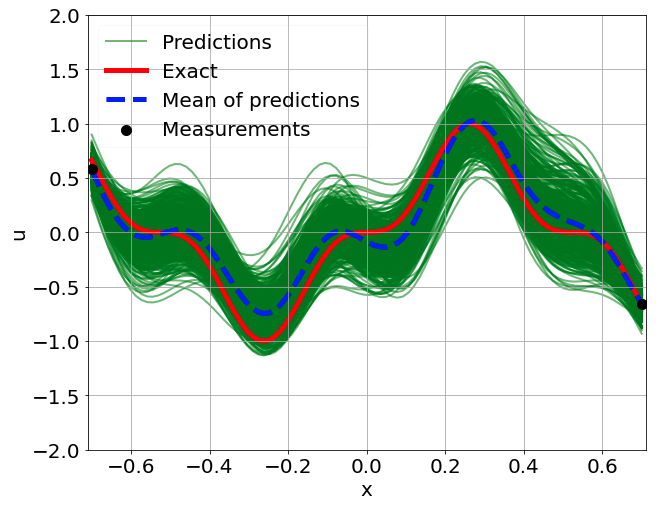} }}%
    \qquad
    \subfloat[\centering Prediction of $f$ with raw solutions]{{\includegraphics[width=7cm]{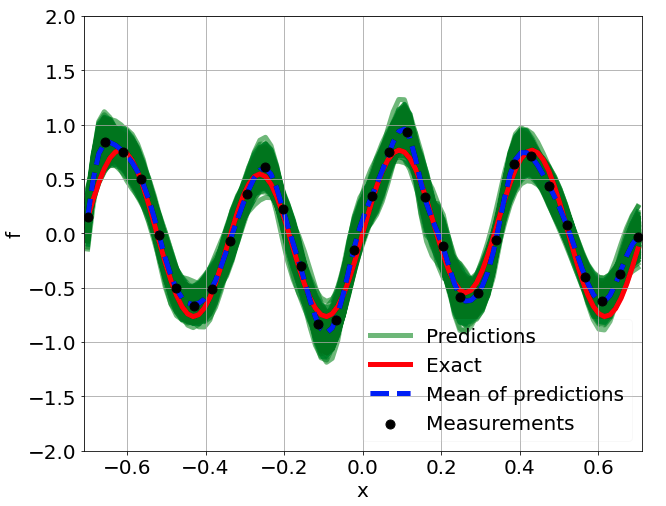} }}%
    \qquad
    \subfloat[\centering Prediction of $u$ with shady uncertain area]{{\includegraphics[width=7cm]{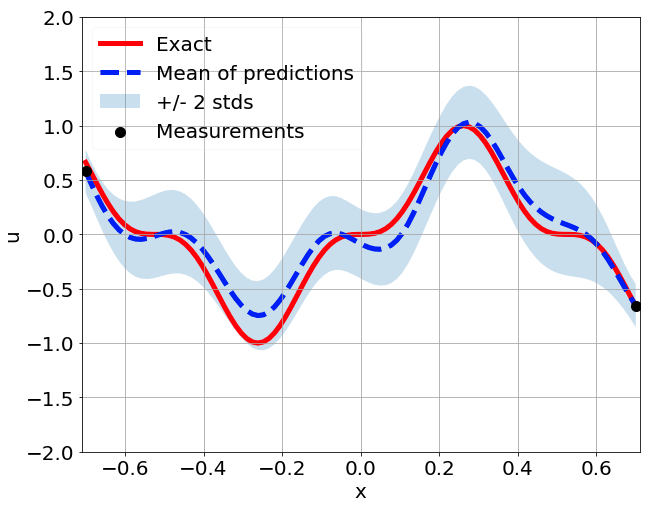} }}%
    \qquad
    \subfloat[\centering Prediction of $f$ with shady uncertain area]{{\includegraphics[width=7cm]{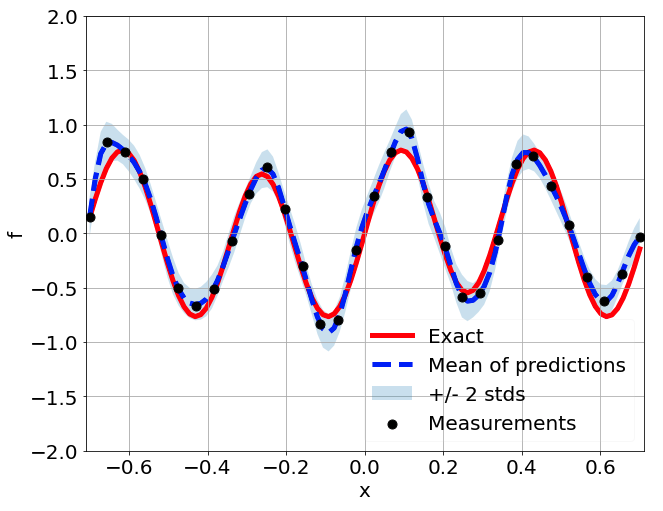} }}%
    \caption{1D nonlinear Poisson equation, predictions of $u$ and source $f$ for case 2. The green lines are the 500 predictions, the dashed 
    blue lines are the mean of the predictions, the red lines are the exact solutions, the light blue shady areas are the mean predictions plus/minus two standard deviations and the black dots are the training data.}
    \label{fig:1d nonlinear possion u and f prediction 0.1 noise}
\end{figure}

The predictions of $u$ and $f$ for the two cases are shown Figure~\ref{fig:1d nonlinear possion u and f prediction 0.01 noise} and Figure~\ref{fig:1d nonlinear possion u and f prediction 0.1 noise}.  Again, we have the same observations about the results as in the last section that exact solution of $u$ is bounded by two standard deviations and the shaded area grows as errors from the measurements increase. Nevertheless, it is shown that the standard deviations of $u$ predictions are smaller than those in the linear Poisson example. This is due to the fact that there are more measurements of $f$ available for training in this example which reduces the uncertainty of predictions.  This observation is also consistent with the results presented in \cite{yang2021b}. 

\subsubsection{2D nonlinear Allen-Cahn equation} \label{2d forward section}
\begin{figure}[H]
    \centering
    \subfloat[\centering Solution $u$ and measurements (green dots)]{{\includegraphics[width=7cm]{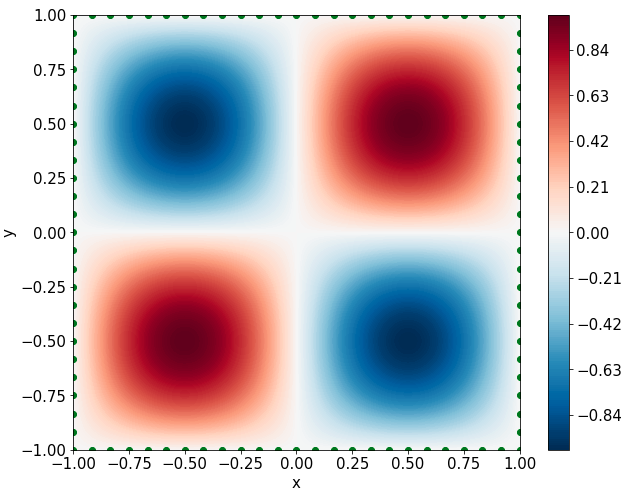} }}%
    \qquad
    \subfloat[\centering Source $f$ and measurements (yellow dots)]{{\includegraphics[width=7cm]{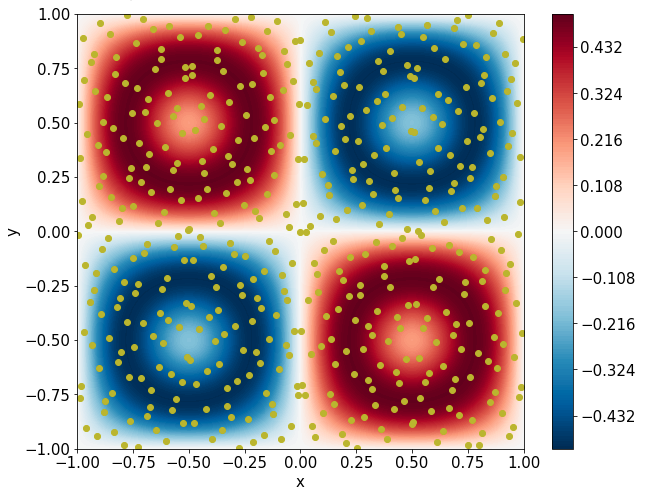} }}%
    \qquad
    \caption{2D nonlinear Allen-Cahn equation, solution and source.}
    \label{fig:2d ac}
\end{figure}

In this section we test the MO-PINN on a two-dimensional problem with the nonlinear Allen-Cahn equation. The equation can be written as
\begin{gather}
    \lambda \left( \frac{\partial^2 u}{\partial x^2} + \frac{\partial^2 u}{\partial y^2} \right) + u\left(u^2-1\right) = f, \qquad x,y \in [-1, 1], \label{2d ac equation} 
\end{gather}
where $\lambda$ is a constant of 0.01 in this example. Again, we adopted the same problem setting as in the paper \cite{yang2021b} to generate the training data, so that we can validate the results and for readers to compare them. The exact solution of $u$ is assumed to be $u=\sin(\pi x)\sin(\pi y)$ and the source $f$ can be calculated with the Equation~\eqref{2d ac equation}. Further, we assume that 500 noisy measurements of $f$ are available in the domain which are randomly distributed. In addition, 25 equally spaced measurements of $u$ are available at each boundary.  Again, we investigate two cases with different noise scales and the errors follow the same distributions as those in \S~\ref{linear case}.  Figure~\ref{fig:2d ac} shows the exact solution $u$ to the problem, the source $f$ and their respective measurements.  The training data is generated in the same fashion as in the one-dimensional cases to provide the outputs of the neural networks, and the goal is also to find the posterior distribution of $u$.

For both cases (small and large noise), the same neural network structure is used for $u$ and $f$, i.e.\ 3 hidden layers consisting of 200 neurons in each followed by the output layer with 2000 neurons.  The ADAM optimizer is used for training with a learning rate of $10^{-3}$ and the results are collected after 50000 epochs. 
\begin{figure}[H]
    \centering
    \subfloat[\centering Prediction of $u$]{{\includegraphics[width=7cm]{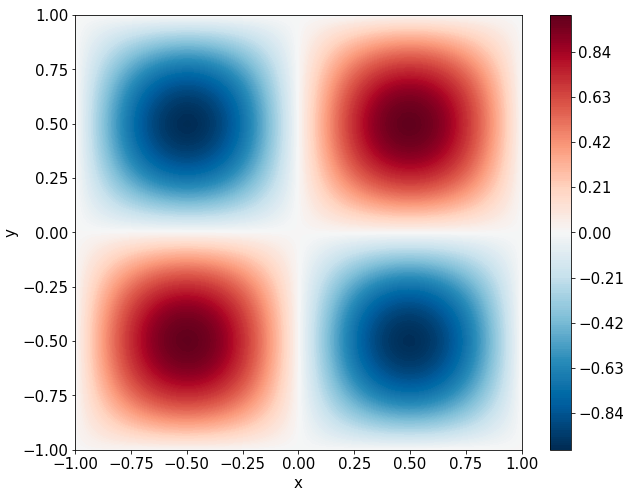} }}%
    \qquad
    \subfloat[\centering Error of $u$ prediction compared to the exact $u$]{{\includegraphics[width=7cm]{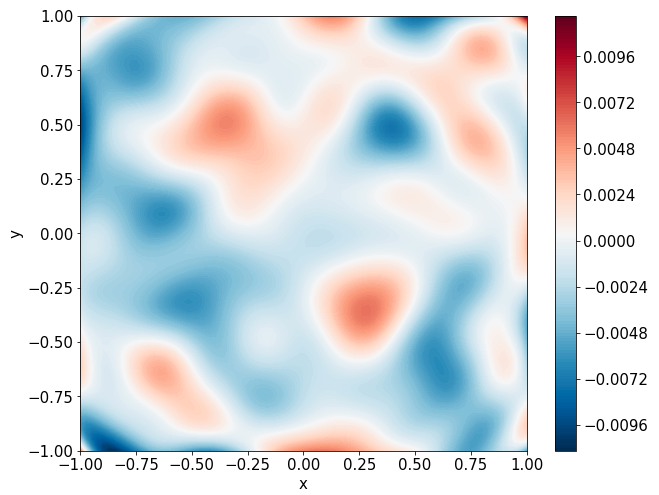} }}%
    \qquad
    \subfloat[\centering Standard deviation of $u$ prediction]{{\includegraphics[width=7cm]{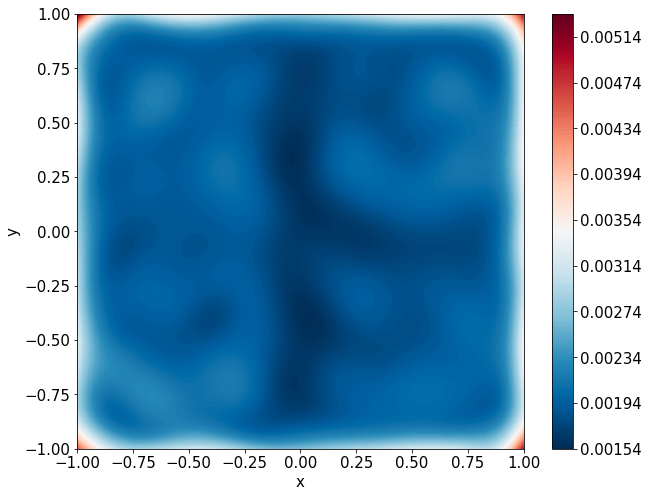} }}%
    \qquad
    \subfloat[\centering Whether the exact $u$ is bounded by 2 stds or not, red region: bounded, blue region: not bounded]{{\includegraphics[width=7cm]{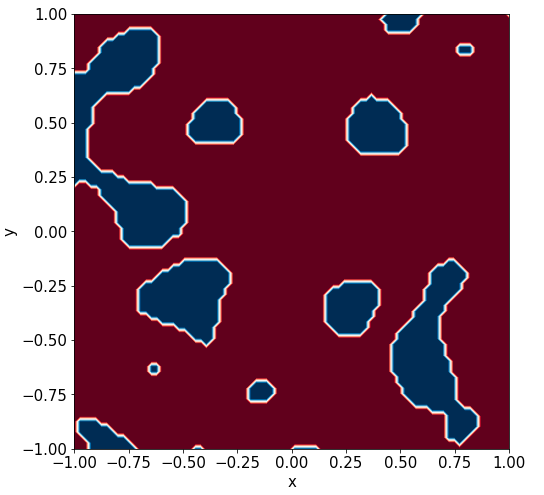} }}%
    \caption{Results of 2D nonlinear Allen-Cahn equation for case 1.}
    \label{fig:2d ac 0.01 noise}
\end{figure}
\begin{figure}[H]
    \centering
    \subfloat[\centering Prediction of $u$]{{\includegraphics[width=7cm]{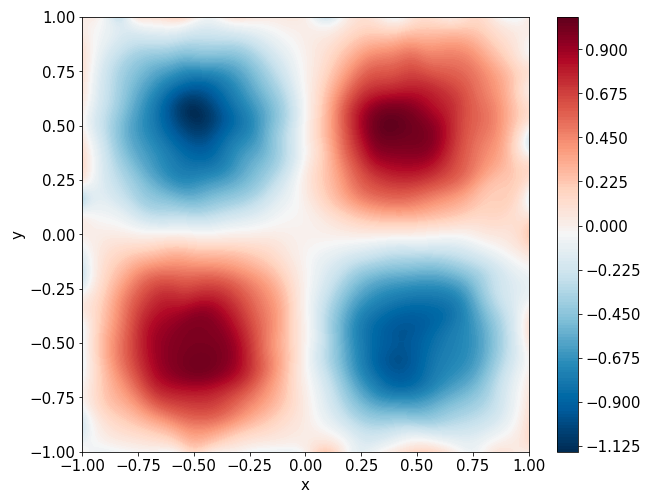} }}%
    \qquad
    \subfloat[\centering Error of $u$ prediction compared to the exact $u$]{{\includegraphics[width=7cm]{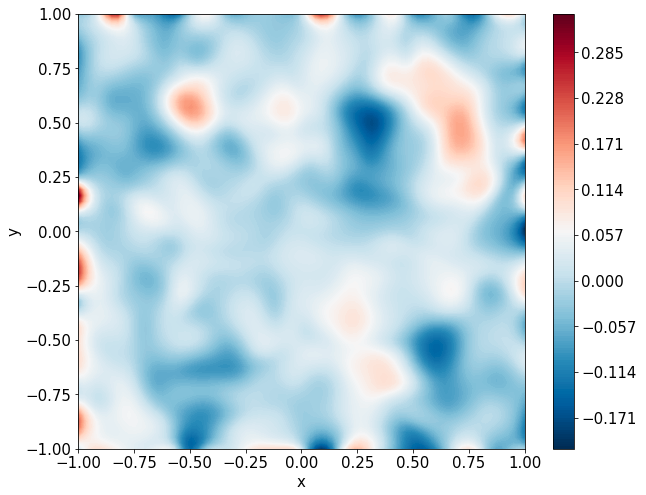} }}%
    \qquad
    \subfloat[\centering Standard deviation of $u$ prediction]{{\includegraphics[width=7cm]{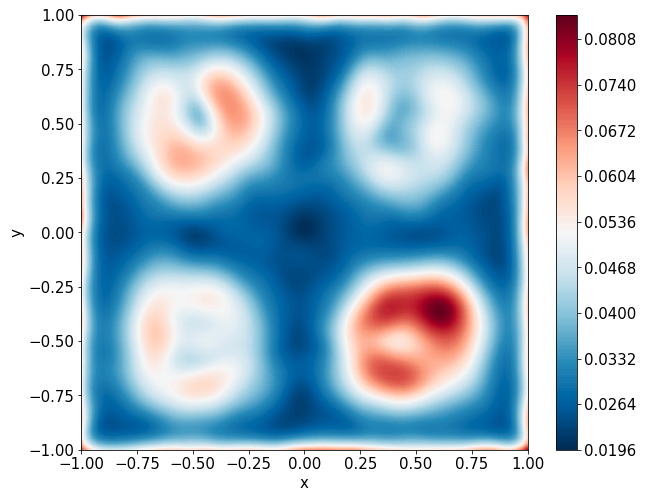} }}%
    \qquad
    \subfloat[\centering Whether the exact $u$ is bounded by 2 stds or not, red region: bounded, blue region: not bounded]{{\includegraphics[width=7cm]{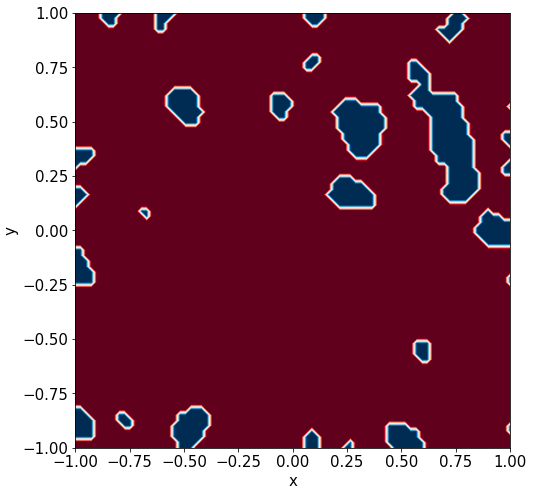} }}%
    \caption{Results of 2D nonlinear Allen-Cahn equation for case 2.}
    \label{fig:2d ac 0.1 noise}
\end{figure}

The results of the two-dimensional forward solution to the Allen-Cahn problem are shown in Figure~\ref{fig:2d ac 0.01 noise} and Figure \ref{fig:2d ac 0.1 noise}.  Similarly, the exact solution to the problem is largely bounded by two standard deviations for both cases with different scales of measurement noise.  In addition, the error between the exact solution and the mean prediction grows as the noise scale increases.

\subsection{Inverse PDE problems}

We define \emph{inverse problems} as those where all or some constitutive parameters are unknown and we are seeking complete solutions to the PDE and the discovery of the unknown parameters.  Examples for linear and nonlinear problems in one and two dimensions are demonstrated.

\subsubsection{1D diffusion-reaction system with nonlinear source term}
Again, we use the same problem as in the paper \cite{yang2021b} to demonstrate the capability of MO-PINN to solve inverse problems. Here, the PDE used in this section is the same as \eqref{1d nonlinear possion equation} except that this time  the coefficient $k$ is assumed to be unknown and the goal is to find the corresponding distribution of $k$ with respect to noisy measurements of $u$ and $f$. We assume that 32 measurements of $f$ are available and equally sampled between $x \in [-0.7, 0.7]$. Additionally, 8 measurements of $u$ are also available and equally sampled in same domain, and the solution of $u$ is assumed to be the same as in \S~\ref{1d nonlinear case}. $k=0.7$ is considered to be the true solution of $k$ and used to generate the data for $f$.   We also test our approach on cases with two noise scales which are the same as in previous sections.  

The neural network structures as well as the training setting remain the same as in the one-dimensional forward problems. In addition to the neural networks, a trainable array is defined for $k$ which corresponds to the multiple outputs of $u_{NN}$ and $f_{NN}$ when formulating the loss function.  The array of $k$ is initialized with random values at the beginning of training, so that it does not require a priori information regarding the $k$ distribution.
\begin{figure}[H]
    \centering
    \subfloat[\centering Prediction of $u$ in case 1]{{\includegraphics[width=7cm]{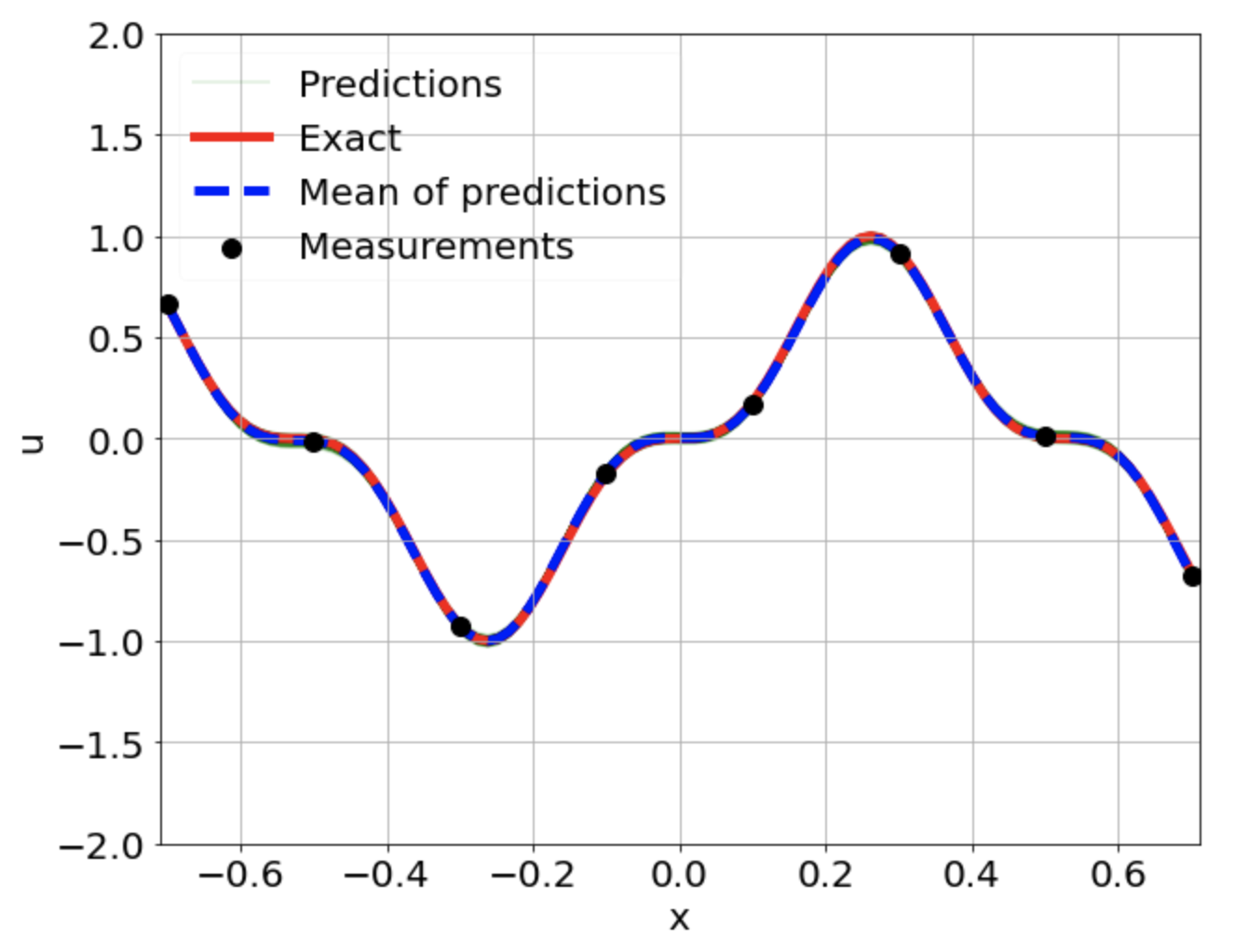} }}%
    \qquad
    \subfloat[\centering Prediction of $u$ in case 2]{{\includegraphics[width=7cm]{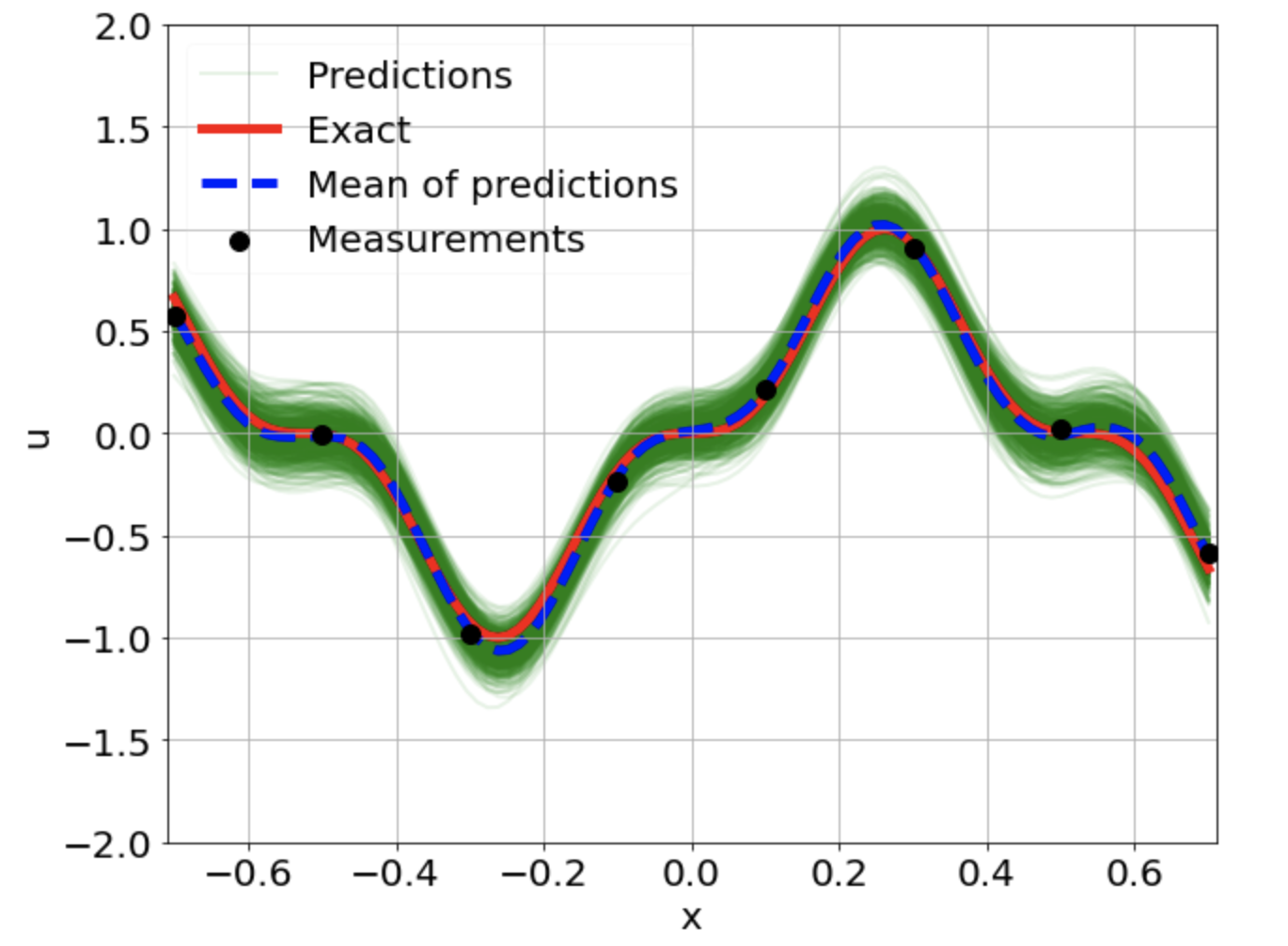} }}%
    \qquad
    \subfloat[\centering Prediction of $f$ in case 1]{{\includegraphics[width=7cm]{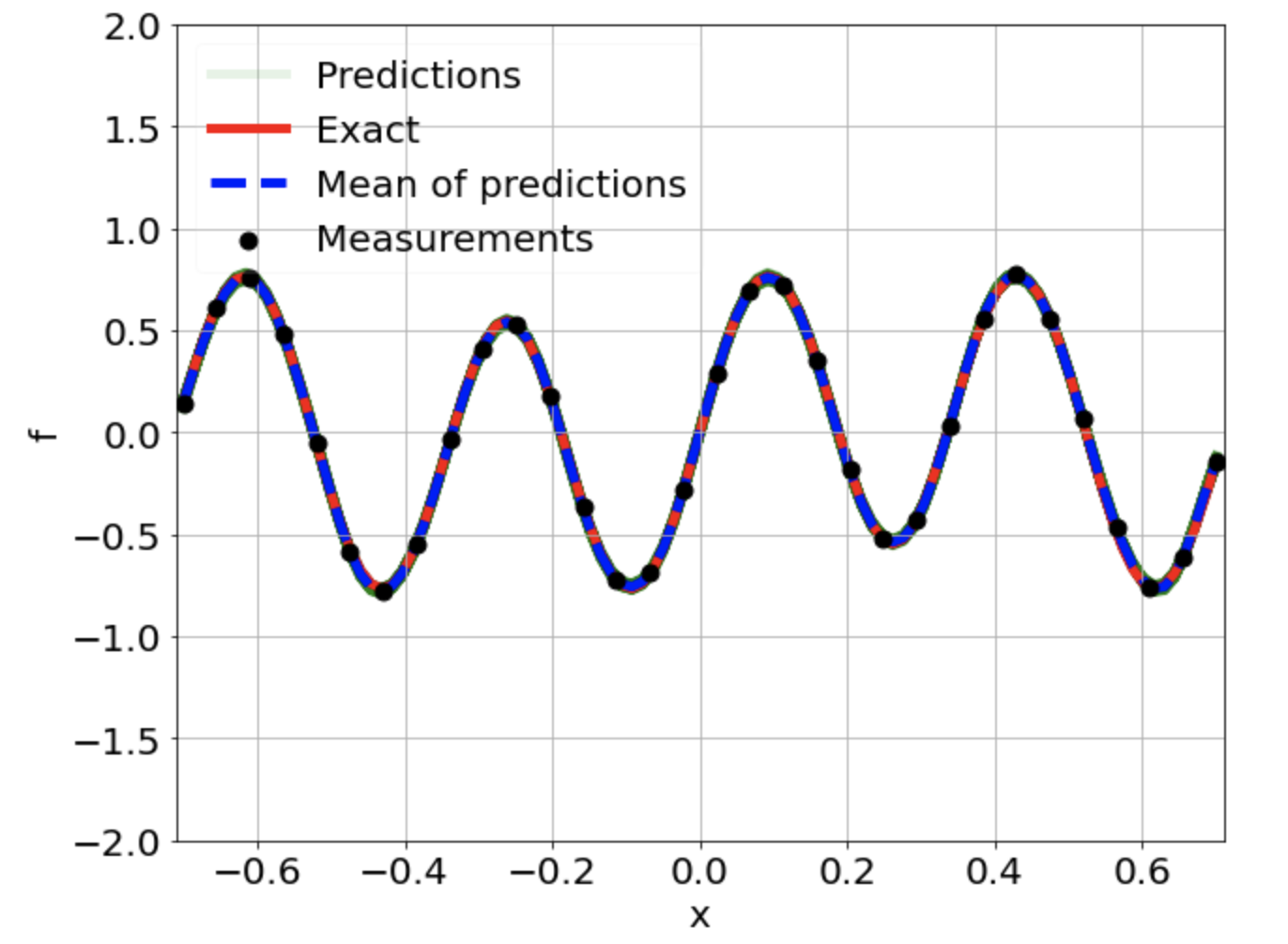} }}%
    \qquad
    \subfloat[\centering Prediction of $f$ in case 2]{{\includegraphics[width=7cm]{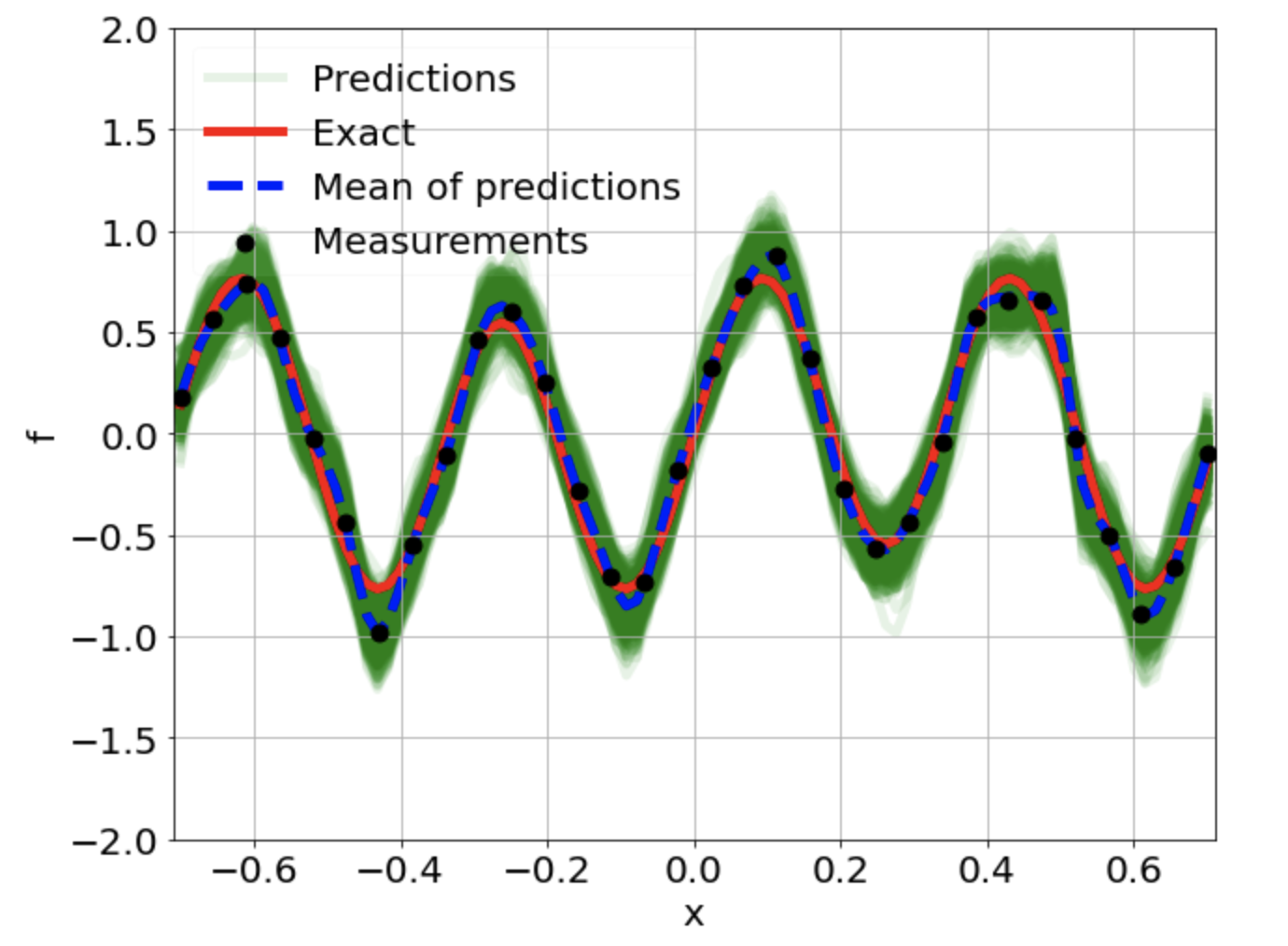} }}%
    \qquad
    \subfloat[\centering Calculated $k$ distribution in case 1 where mean is 0.698 and standard deviation is 0.006]{{\includegraphics[width=7cm]{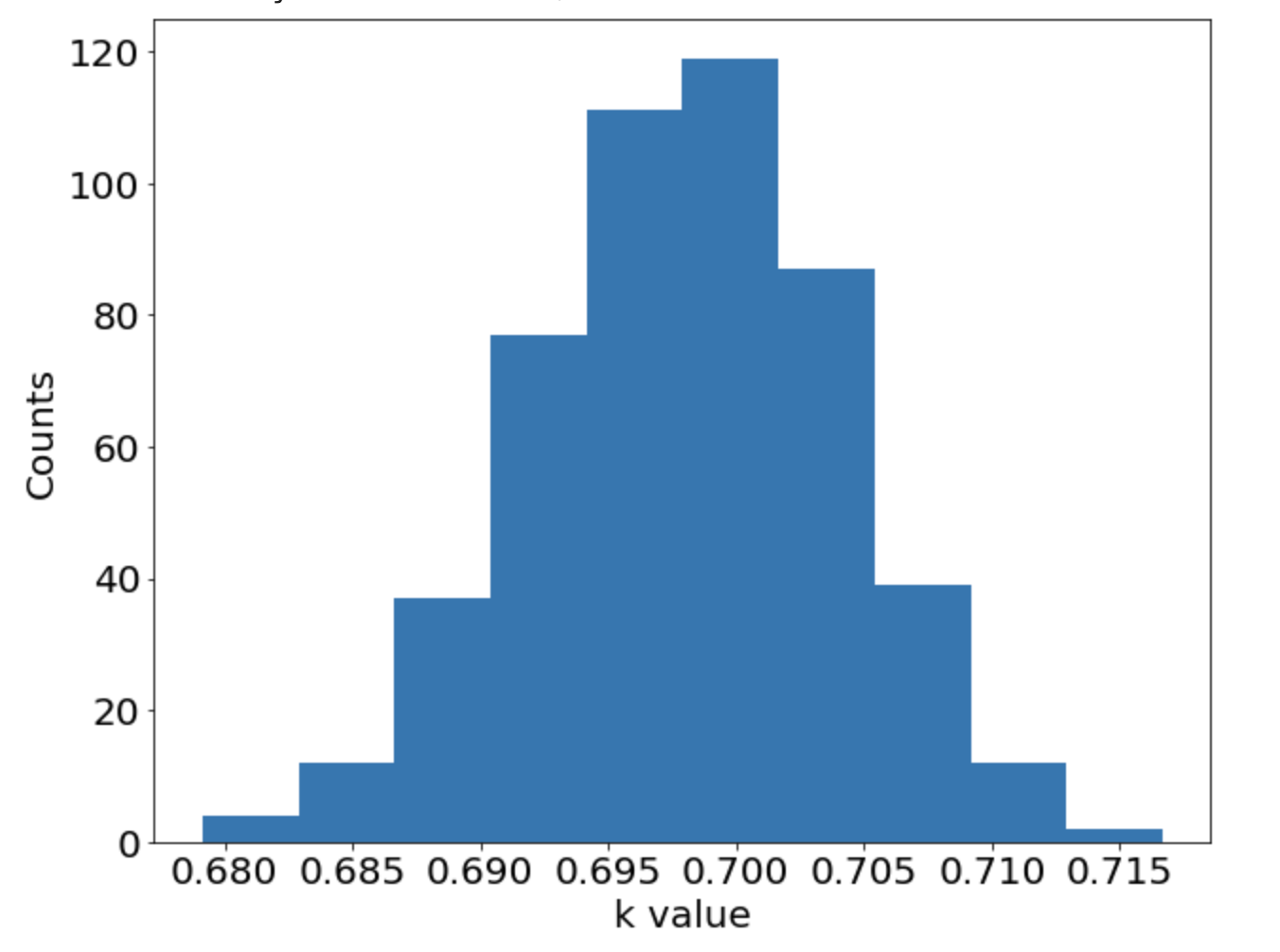} }}%
    \qquad
    \subfloat[\centering Calculated $k$ distribution in case 2 where mean is 0.678 and standard deviation is 0.063]{{\includegraphics[width=7cm]{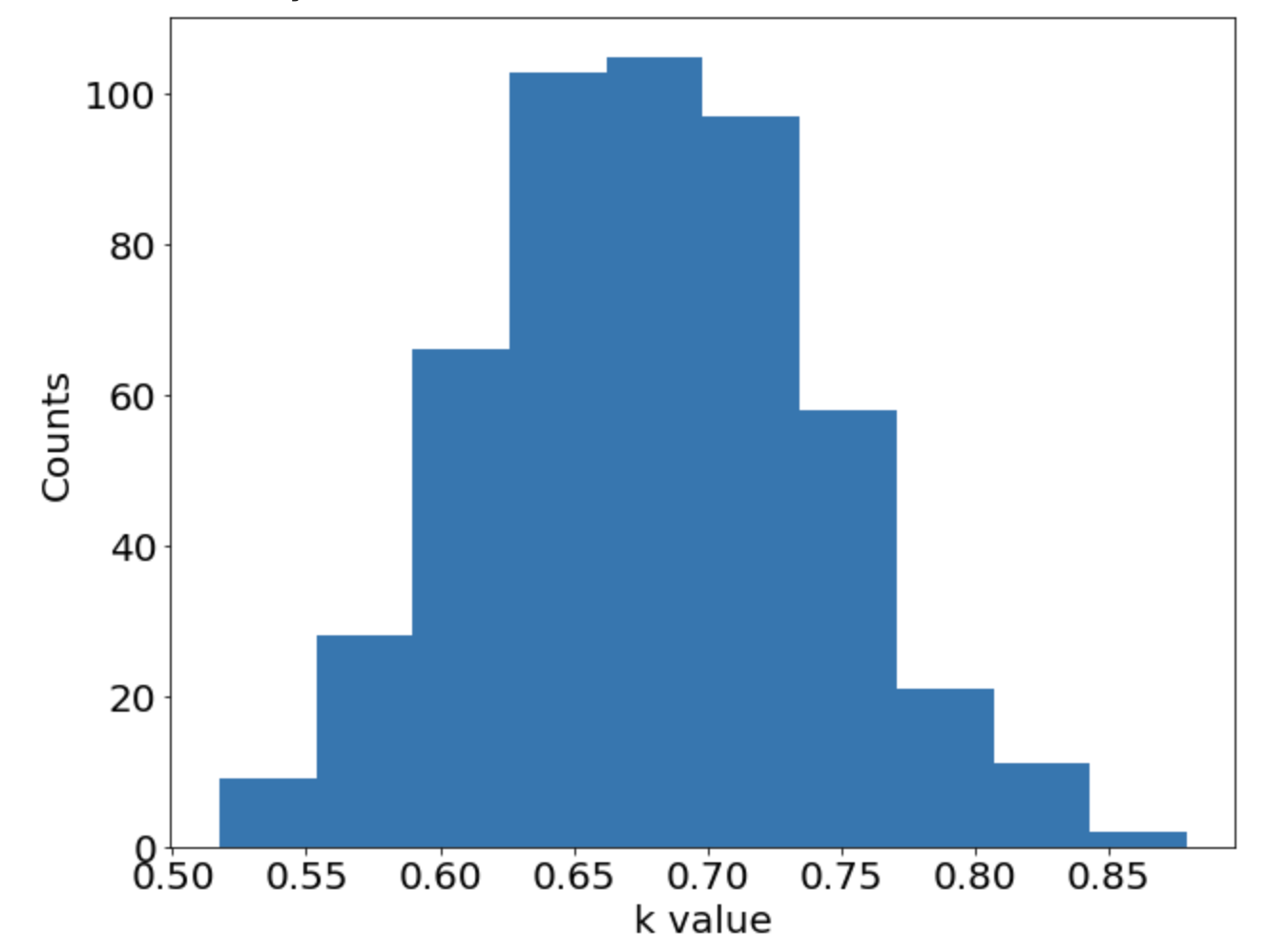} }}%
    \caption{Results of 1D inverse nonlinear diffusion-reaction problem for both cases.}
    \label{fig:1d inverse}
\end{figure}

A summary of results of this one-dimensional inverse problem for both noise scale cases is shown in Figure~\ref{fig:1d inverse}.  Similar to the results in the forward problems, the mean predictions of $u$ and $f$ pass through the noisy measurements.   In addition, we end up with distributions of $k$ which is the objective of the inverse problem training. It is shown that the values in the array of $k$ follow a normal distribution as expected. Moreover, the true solution $k=0.7$ is bounded by the mean prediction of $k$ plus/minus two standard deviations which indicates that MO-PINN is capable of returning a reasonable posterior distribution in this inverse problem. 
\begin{table}[h!]
    \caption{Mean and standard deviation of predicted $k$ of 6 simulations with different random seeds for case 2.}
    \begin{center}
      \begin{tabular}{l|c|r} 
        \textbf{No.} & \textbf{Mean} &  \textbf{Std}\\
        \hline
        1 & $0.673$ & 0.073  \\
        2 & $0.681$ & 0.075  \\
        3 & $0.675$ & 0.072  \\
        4 & $0.676$ & 0.073  \\
        5 & $0.674$ & 0.071  \\
        6 & $0.675$ & 0.073 \\
      \end{tabular}
      \label{tab:1d inverse consistency}
    \end{center}
\end{table}

We also tested the consistency of our solutions by training on the same measurements but different random seeds to initialize the neural networks.  Results of 6 simulations are shown in Table~\ref{tab:1d inverse consistency}, and it demonstrates that all return all of the distributions of predicted $k$ are in agreement.  

\begin{figure}[H]
    \centering
    \subfloat[\centering 10 outputs, mean is 0.670 and std is 0.066]{{\includegraphics[width=7cm]{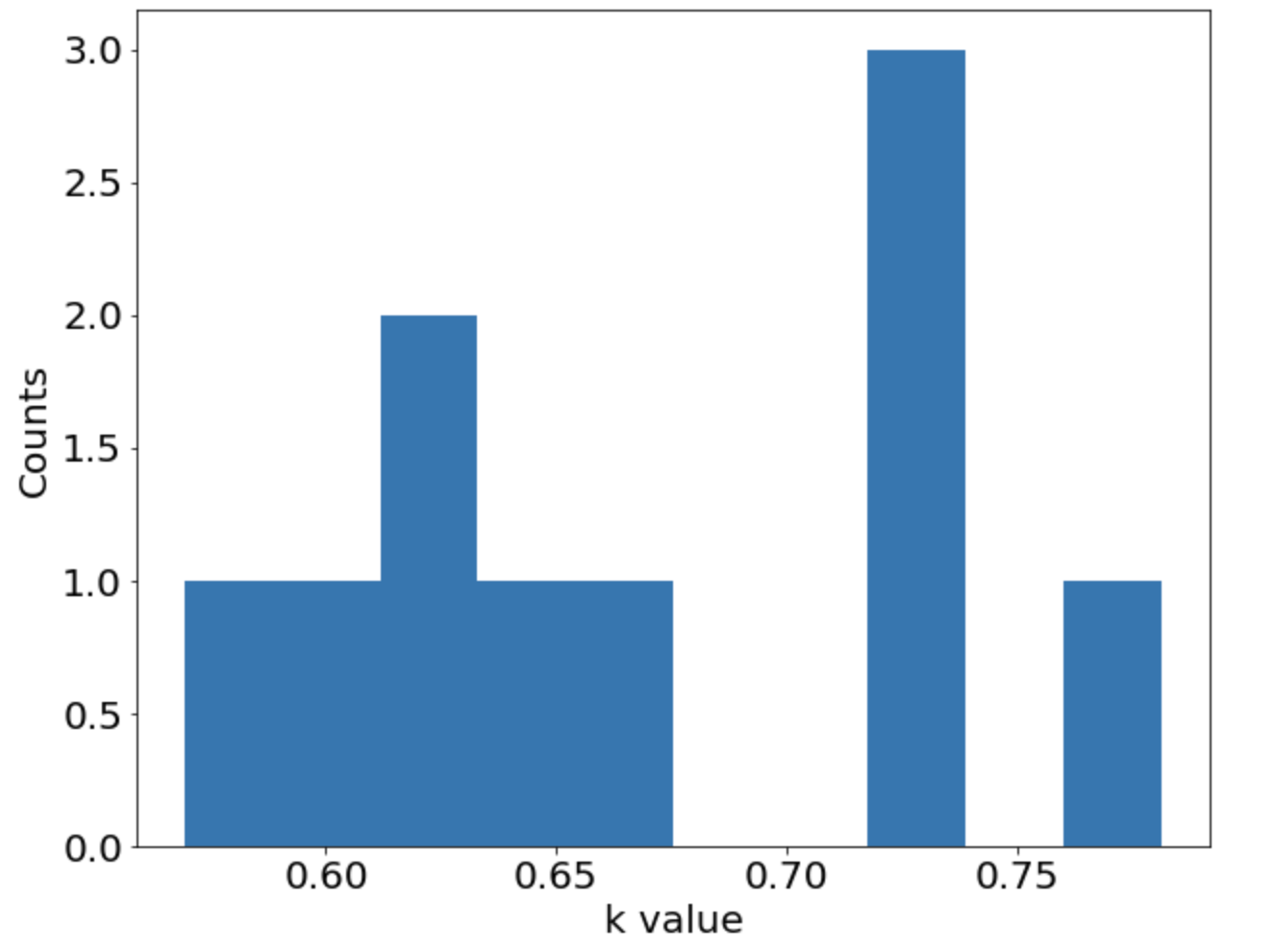} }}%
    \qquad
    \subfloat[\centering 50 outputs, mean is 0.684 and std is 0.063]{{\includegraphics[width=7cm]{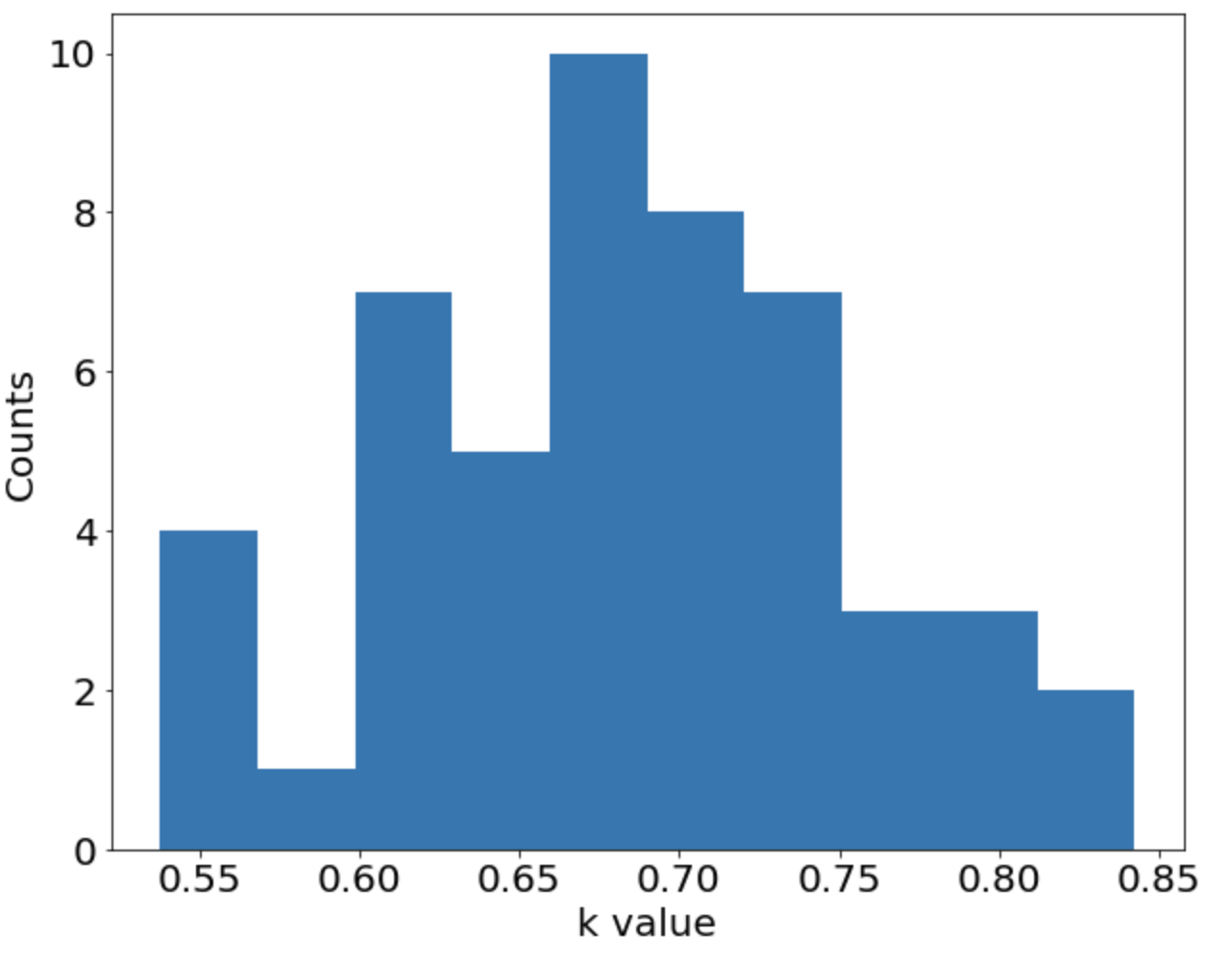} }}%
    \qquad
    \subfloat[\centering 100 outputs, mean is 0.668 and std is 0.074]{{\includegraphics[width=7cm]{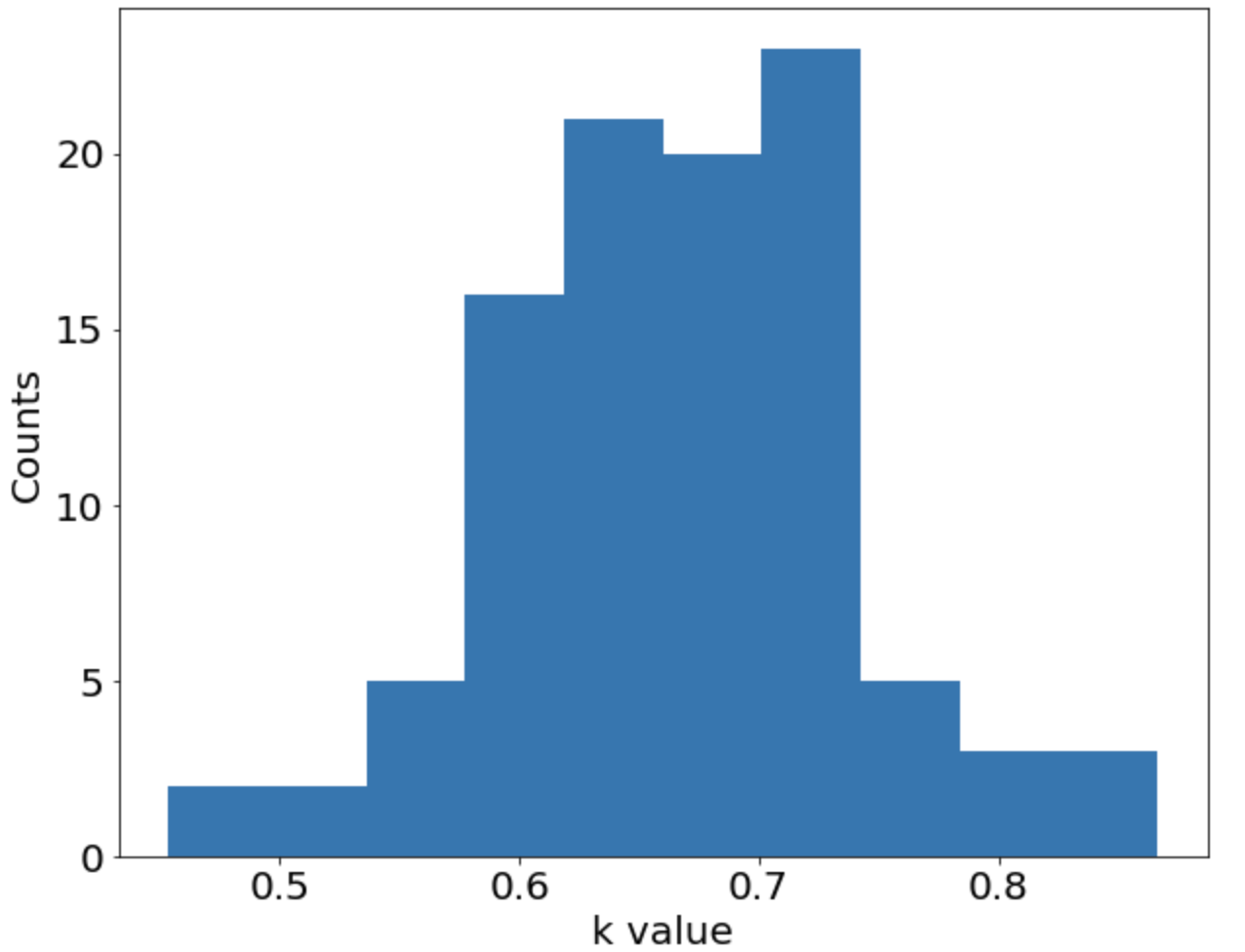} }}%
    \qquad
    \subfloat[\centering 500 outputs, mean is 0.673 and std is 0.072]{{\includegraphics[width=7cm]{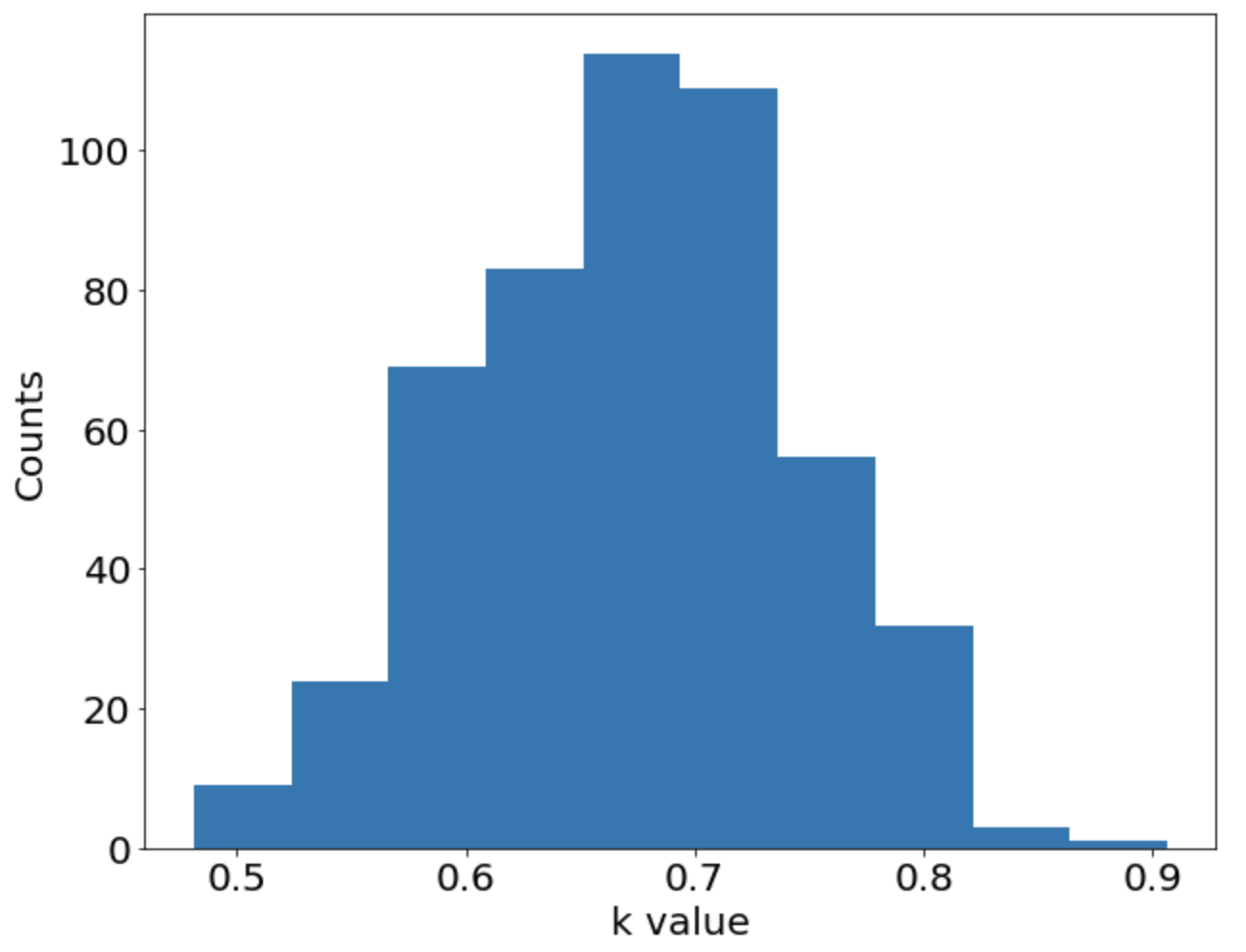} }}%
    \caption{Distributions of predicted $k$ with different number of outputs for case 2.}
    \label{fig:1d inverse convergence}
\end{figure}

Additionally, we solved the problem with different number of outputs for the neural networks and the respective $k$ distributions are shown in Figure~\ref{fig:1d inverse convergence}.
It gives us a better intuition of the process that the posterior distribution forms as the number of outputs increases. 

\subsubsection{Two-dimensional nonlinear diffusion-reaction system}
For the two-dimensional inverse problem, we consider the following PDE,
\begin{gather}
    \lambda \left(\frac{\partial^2 u}{\partial x^2} + \frac{\partial^2 u}{\partial y^2}\right) + k u^2 = f, \qquad x,y \in [-1, 1], \label{2d inverse equation} 
\end{gather}
where the constant $\lambda=0.01$ and $k$ is an unknown parameter to be solved for with noisy measurements of $u$ and $f$. The same exact solution of $u$ as in \S~\ref{2d forward section} is considered here, that $u=\sin(\pi x)\sin(\pi y)$.  One hundred noisy measurements of $u$ and $f$ are assumed to be available across the domain. In addition, 25 equally spaced measurements of $u$ are also available at each boundary.  $k=1$ is the exact solution in this example and used to generate the training data.  We also test the MO-PINN on two cases with the different noise scales. The errors in the measurements follow the same distributions as in previous sections.  The structure of neural networks and the hyperparameters for training are kept the same as in \S~\ref{2d forward section}.  The only difference is that we have an additional trainable array for $k$ which is of the size 2000 corresponding to the number of outputs for the two neural networks. 
\begin{figure}[H]
    \centering
    \subfloat[\centering Solution $u$ and measurements (green dots)]{{\includegraphics[width=7cm]{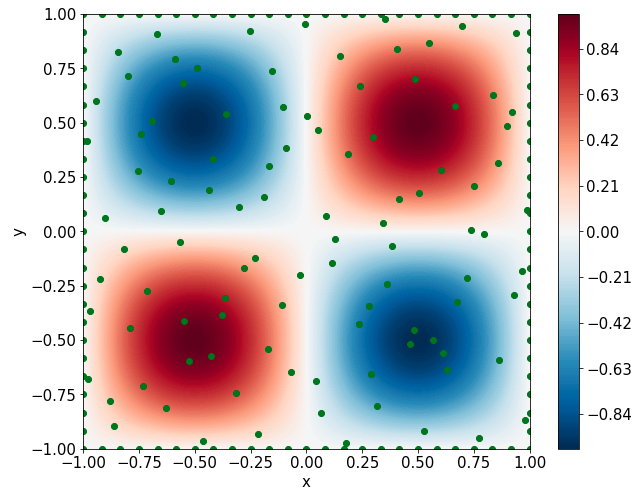} }}%
    \qquad
    \subfloat[\centering Source $f$ and measurements (yellow dots)]{{\includegraphics[width=7cm]{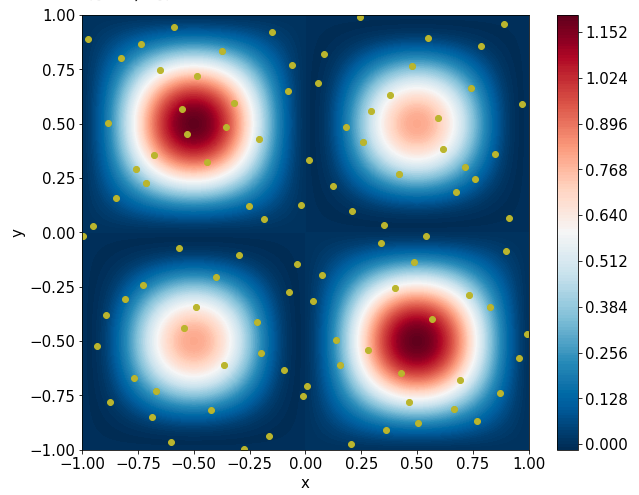} }}%
    \qquad
    \caption{2D nonlinear diffusion-reaction equation, solution and source.}
    \label{fig:2d inverse}
\end{figure}
\begin{figure}[H]
    \centering
    \subfloat[\centering k distribution for case 1, mean is 0.995 and std is 0.005]{{\includegraphics[width=7cm]{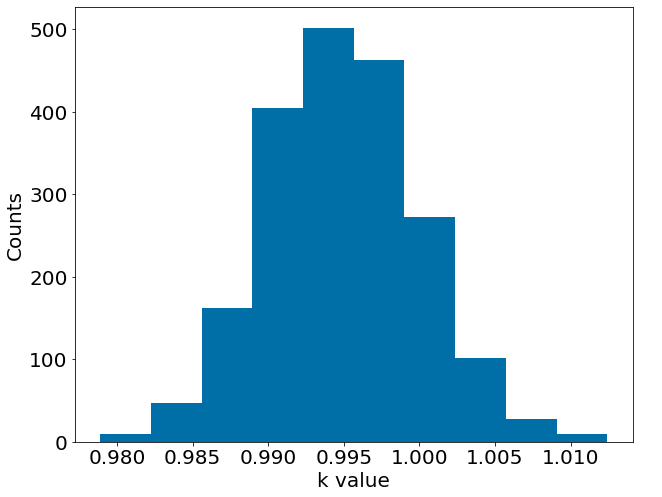} }}%
    \qquad
    \subfloat[\centering k distribution for case 2, mean is 1.02 and std is 0.05]{{\includegraphics[width=7cm]{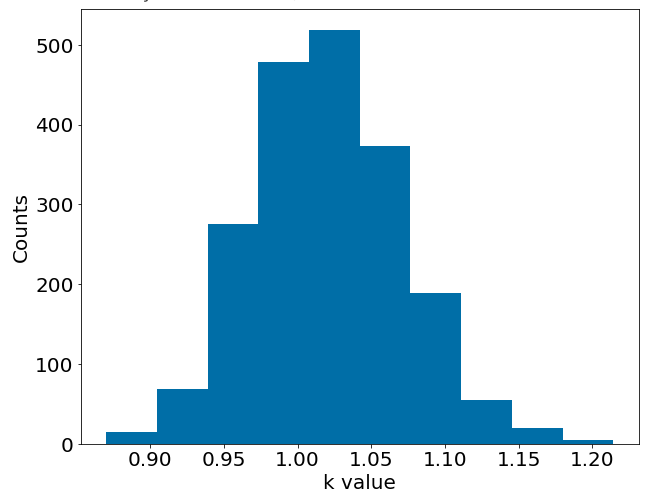} }}%
    \qquad
    \caption{2D nonlinear diffusion-reaction equation, solution and source.}
    \label{fig:2d inverse k}
\end{figure}

The resulting $k$ distributions for both cases are shown in Figure~\ref{fig:2d inverse k}. Again, we have similar observations as in the previous section that  the exact solution of $k$ is within the two standard deviations confidence interval. Also, as the error in the measurements grow, the standard deviation of the $k$ distribution increases. 

\subsection{Incorporation of additional prior knowledge to improve the prediction}
In this section, we will show that the MO-PINN approach is capable of taking prior knowledge from different sources into consideration to improve the predictions.  We take the problem in \S~\ref{linear case} with the large noise scale as example for demonstration.  First, we solve the same problem with the finite element method using a Monte Carlo approach.  This is a common way to infer a distribution with only forward models available. In this work, the FEniCS package \cite{alnaes2015fenics} is used as the forward solver.  In order to solve the problem with the FEM, the full information of source $f$ needs to be provided or at least at all quadrature points. Here, we made a simple assumption that the source $f$ is a linear function between measurements for the FEM approach. Therefore, for each set of training data of $u$ and $f$ in Section~\ref{linear case}, there is a corresponding FEM solution to the problem, and they form the final posterior distributions as we did before.  This can also be seen as another validation of the MO-PINN approach in solving this 1D problem. 
\begin{figure}[H]
    \centering
    \subfloat[\centering Prediction of u with MO-PINN]{{\includegraphics[width=7cm]{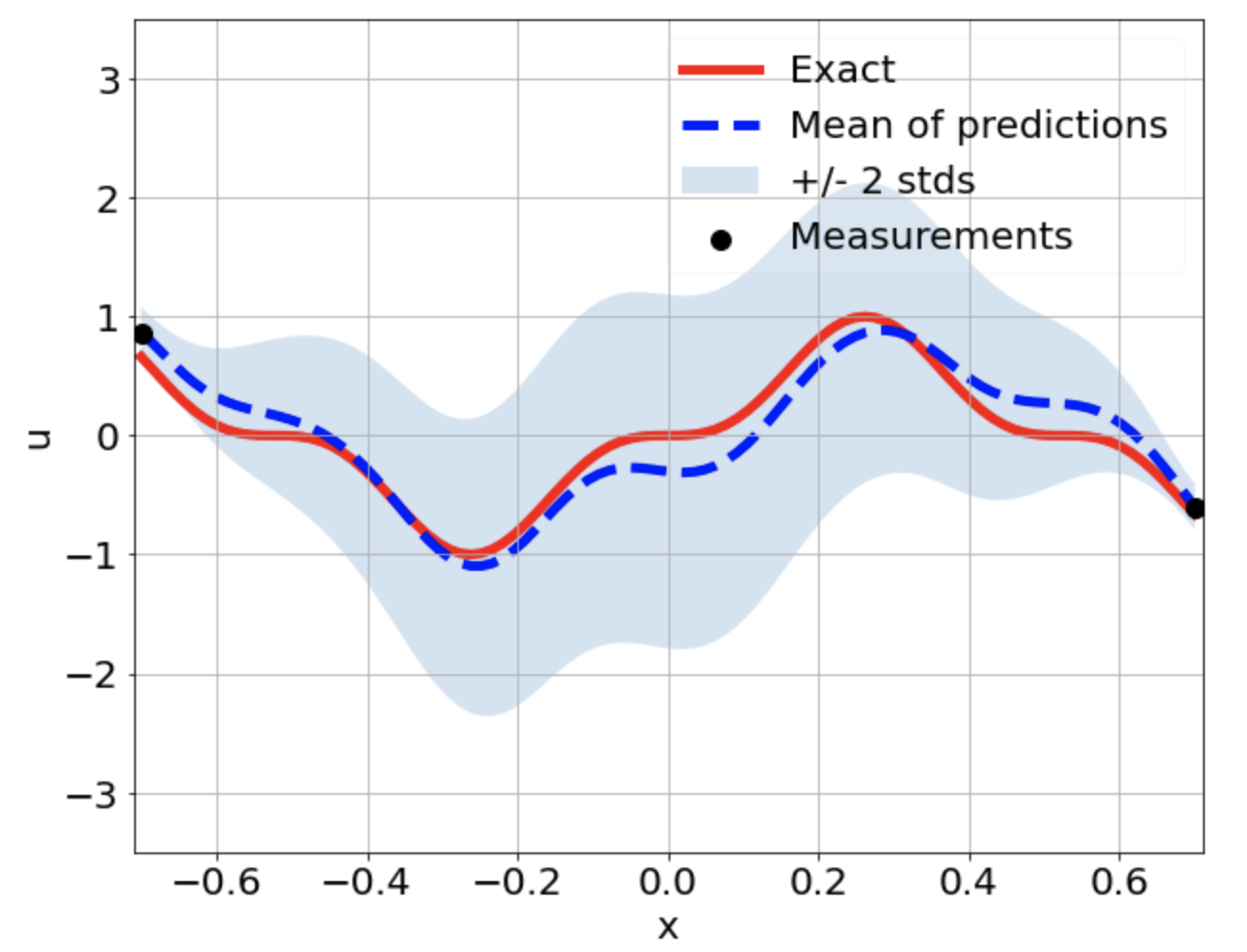} }}%
    \qquad
    \subfloat[\centering Prediction of u with FEniCS]{{\includegraphics[width=7cm]{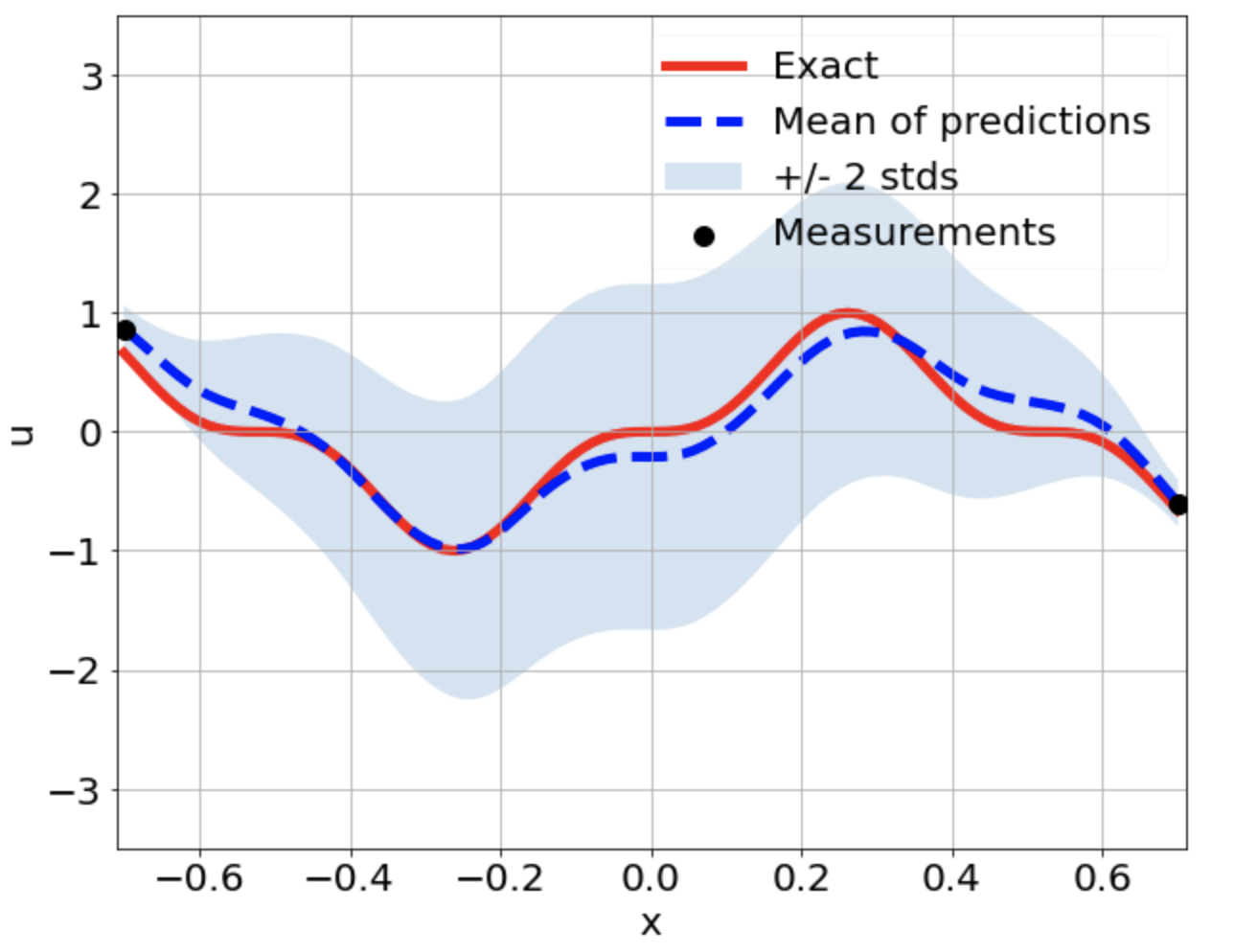} }}%
    \qquad
    \subfloat[\centering Prediction of f with MO-PINN]{{\includegraphics[width=7cm]{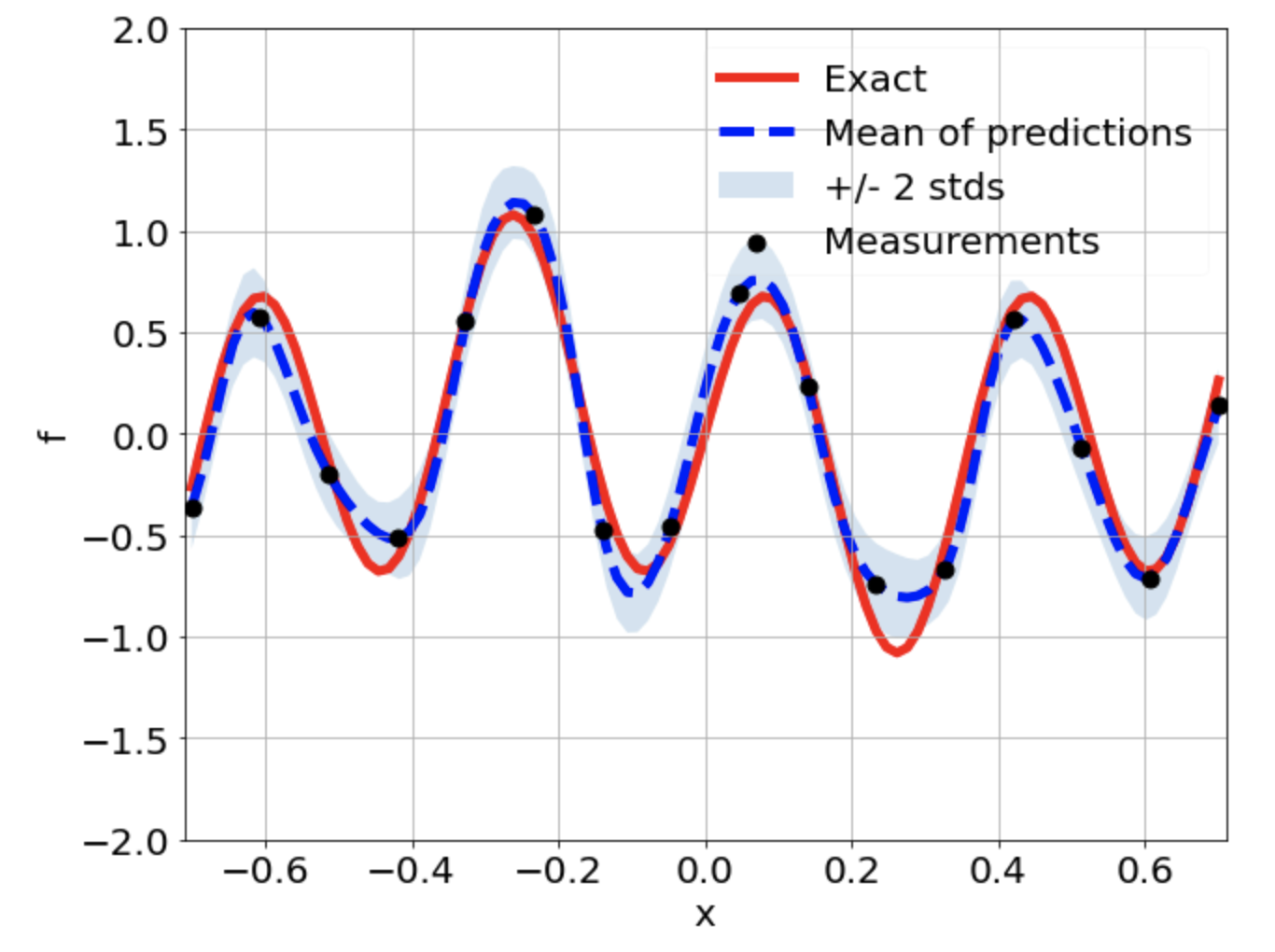} }}%
    \qquad
    \subfloat[\centering f under the linear interpolation assumption for FEniCS]{{\includegraphics[width=7cm]{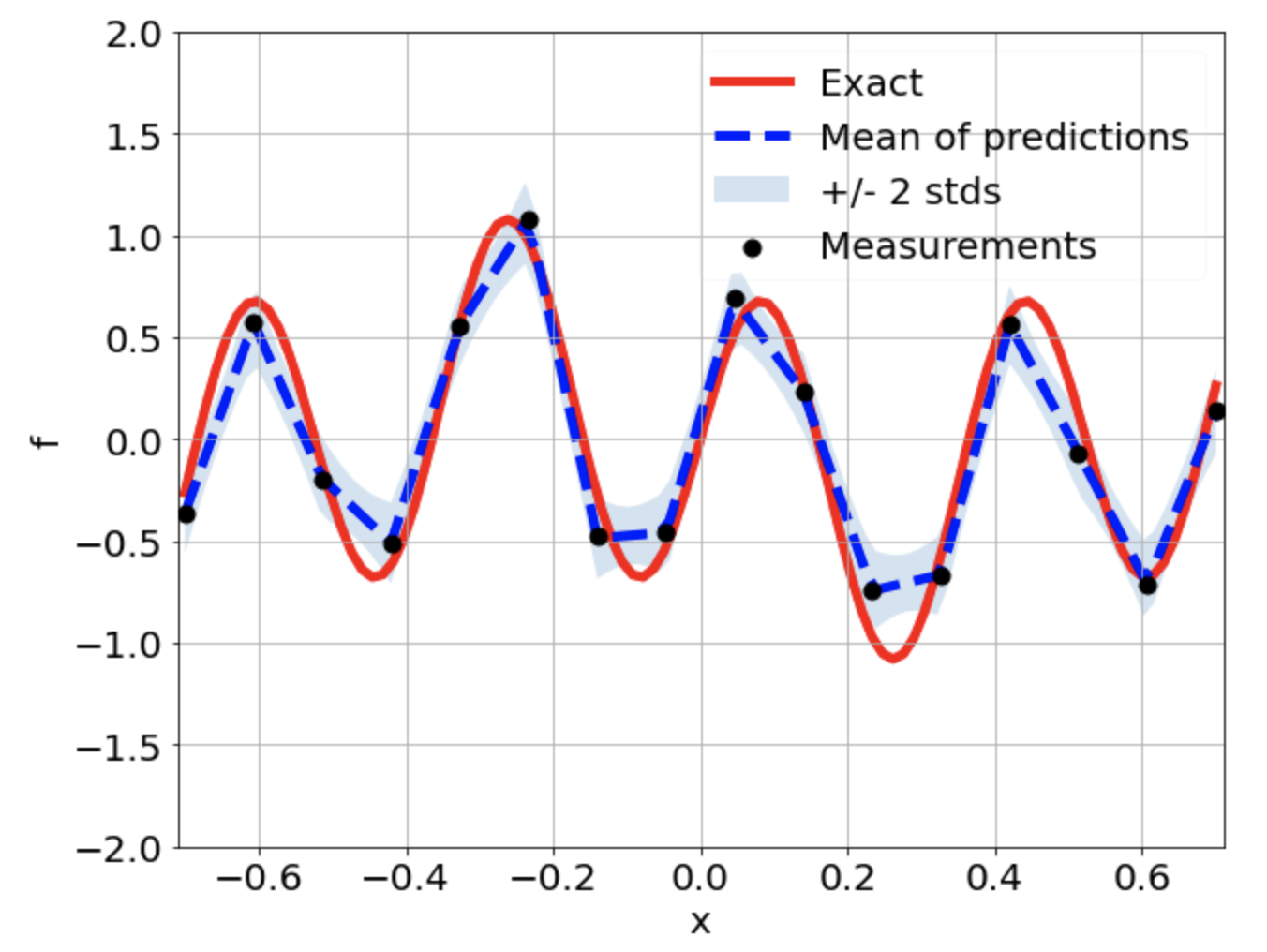} }}%
    \caption{Comparison between MO-PINN and FEM Monte Carlo approach using the same measurements.}
    \label{fig:fenics comp}
\end{figure}

In Figure~\ref{fig:fenics comp}, it shows the comparison of results between MO-PINN and the FEM Monte Carlo approach. The same number of forward simulations (500) were run to generate the posterior distributions for comparison.

\begin{figure}[H]
    \centering
    \subfloat{{\includegraphics[width=10cm]{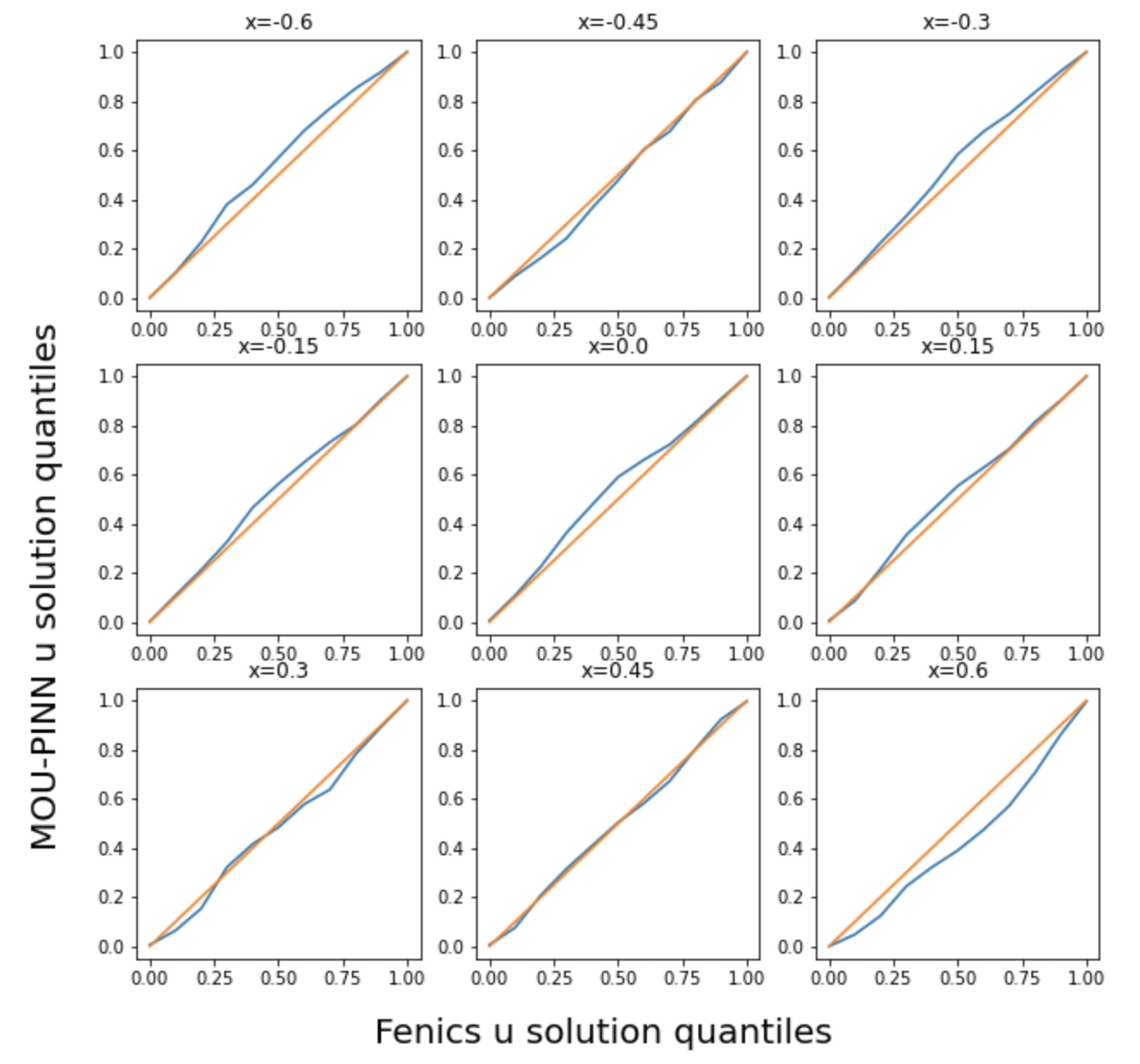} }}%
    \caption{Quantile-quantile plot of the $u$ prediction at 9 locations.}
    \label{fig:quantile}
\end{figure}

The resulting distributions of $u$ are almost the same and the minor differences are caused by different assumptions on $f$ in each approach. In MO-PINN, the neural networks naturally returns a smooth function interpolation of the measurements while we enforce $f$ to be linear in between for the FEM simulation.  In Figure~\ref{fig:quantile}, it shows another comparison of the two distributions of $u$ with a quantile-quantile plot.  This type of plot is used to answer the question, ``What percentage of the data lies in the $x$ quantile of the prediction distribution?" where, $x$ is $10\%, 20\%, \ldots, 100\%$.

\begin{figure}[H]
    \centering
    \subfloat[\centering Prediction of u with only measurements]{{\includegraphics[width=7cm]{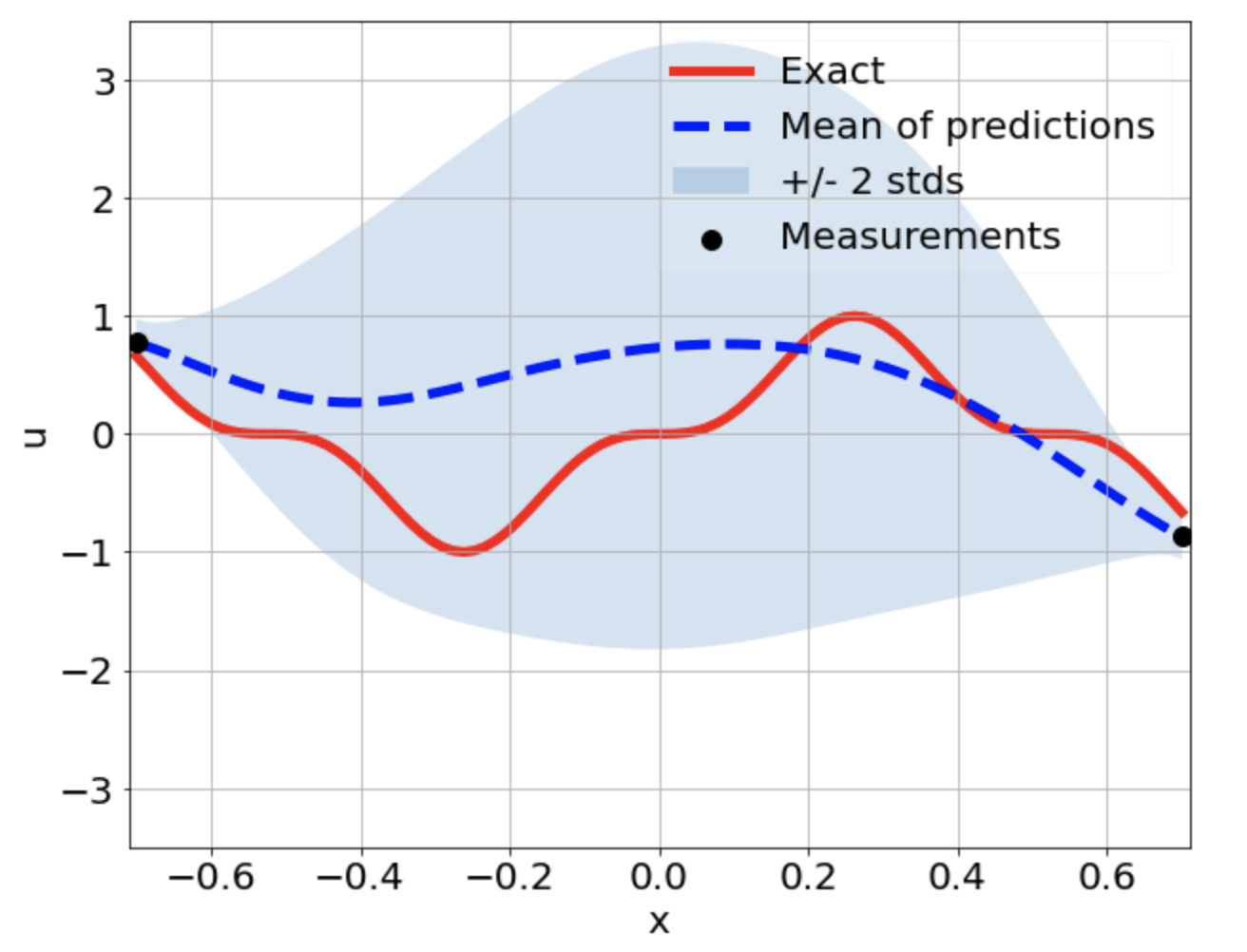} }}%
    \qquad
    \subfloat[\centering Prediction of f with only measurements]{{\includegraphics[width=7cm]{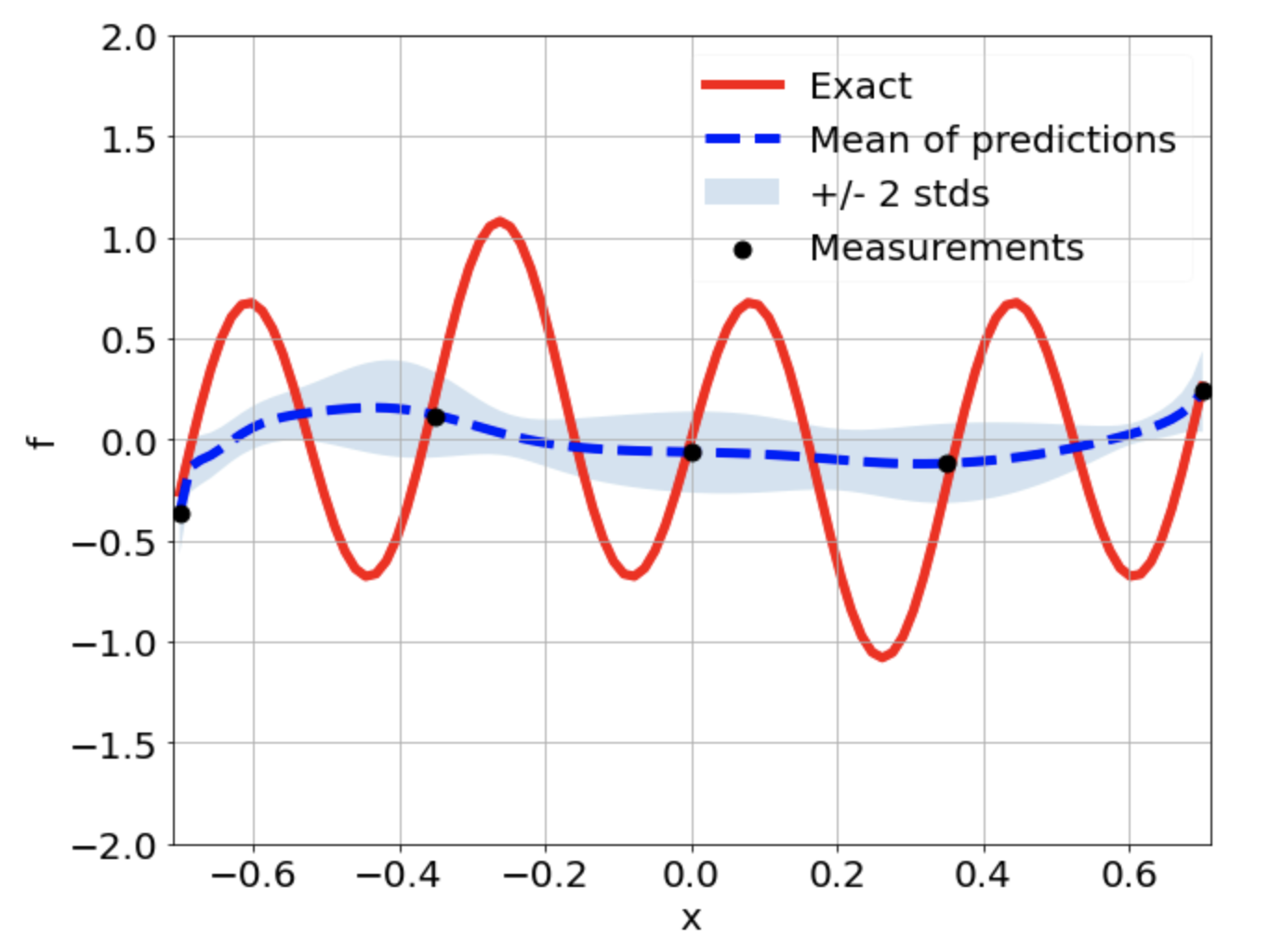} }}%
    \qquad
    \subfloat[\centering Prediction of u with measurements and means]{{\includegraphics[width=7cm]{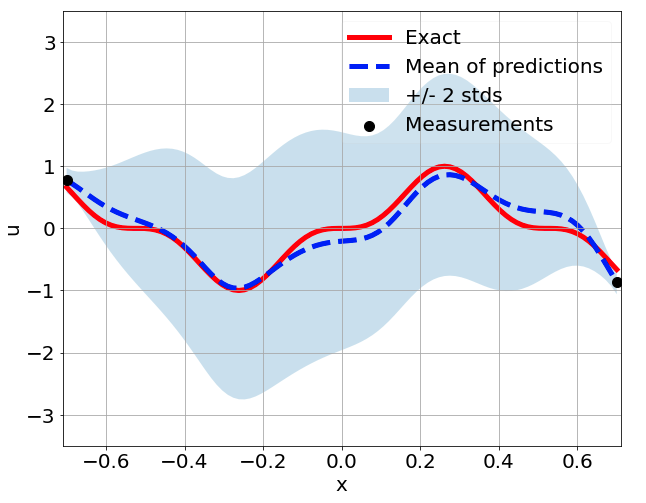} }}%
    \qquad
    \subfloat[\centering Prediction of f with measurements and means]{{\includegraphics[width=7cm]{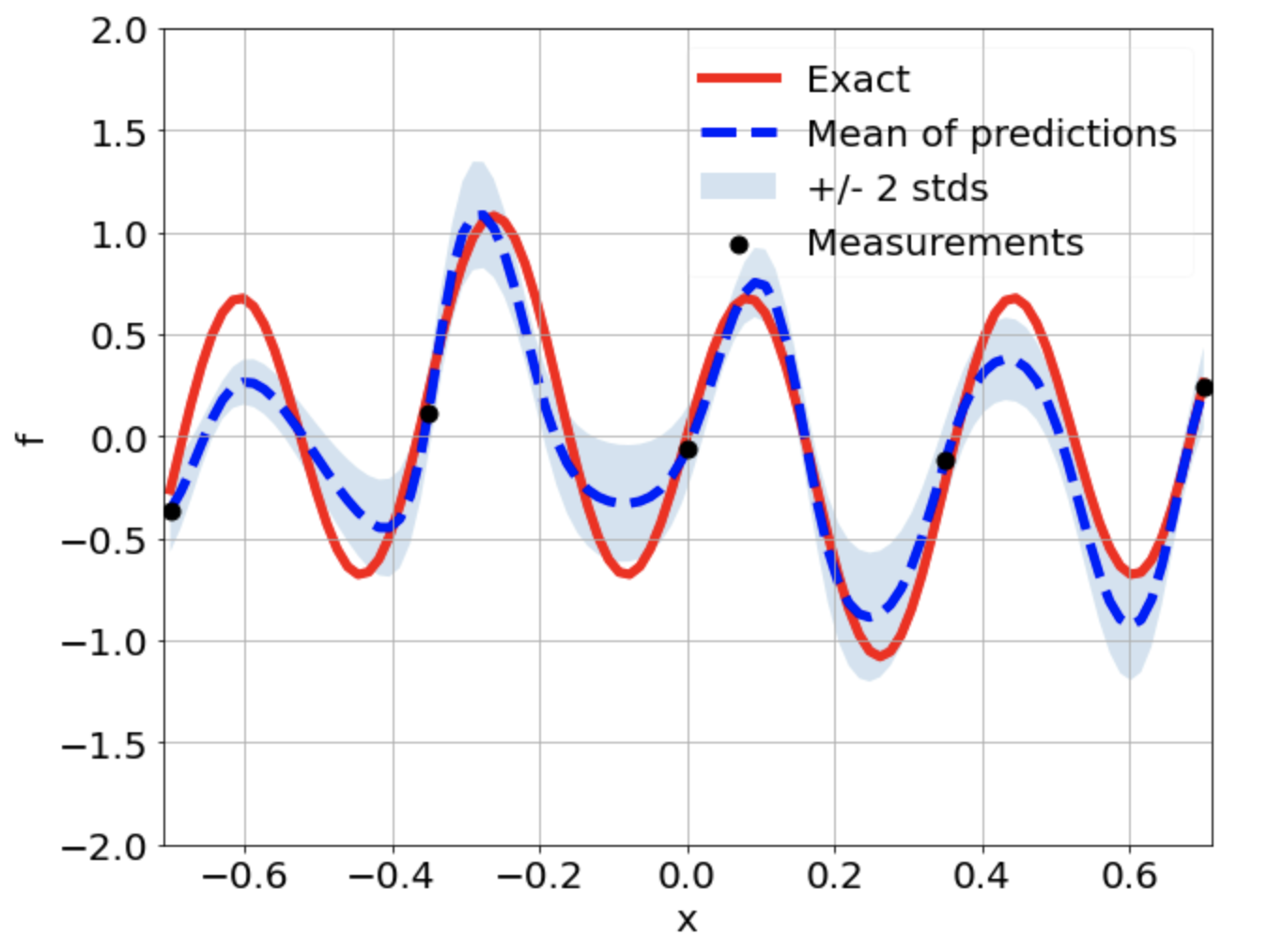} }}%
    \qquad
    \subfloat[\centering Prediction of u with measurements, means and stds]{{\includegraphics[width=7cm]{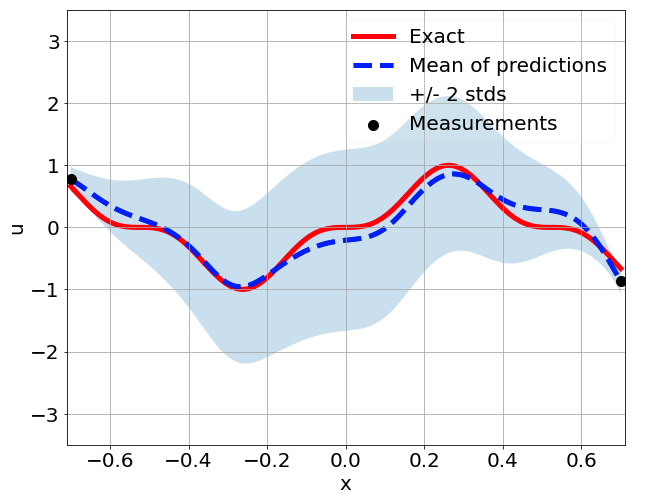} }}%
    \qquad
    \subfloat[\centering Prediction of f with measurements, means and stds]{{\includegraphics[width=7cm]{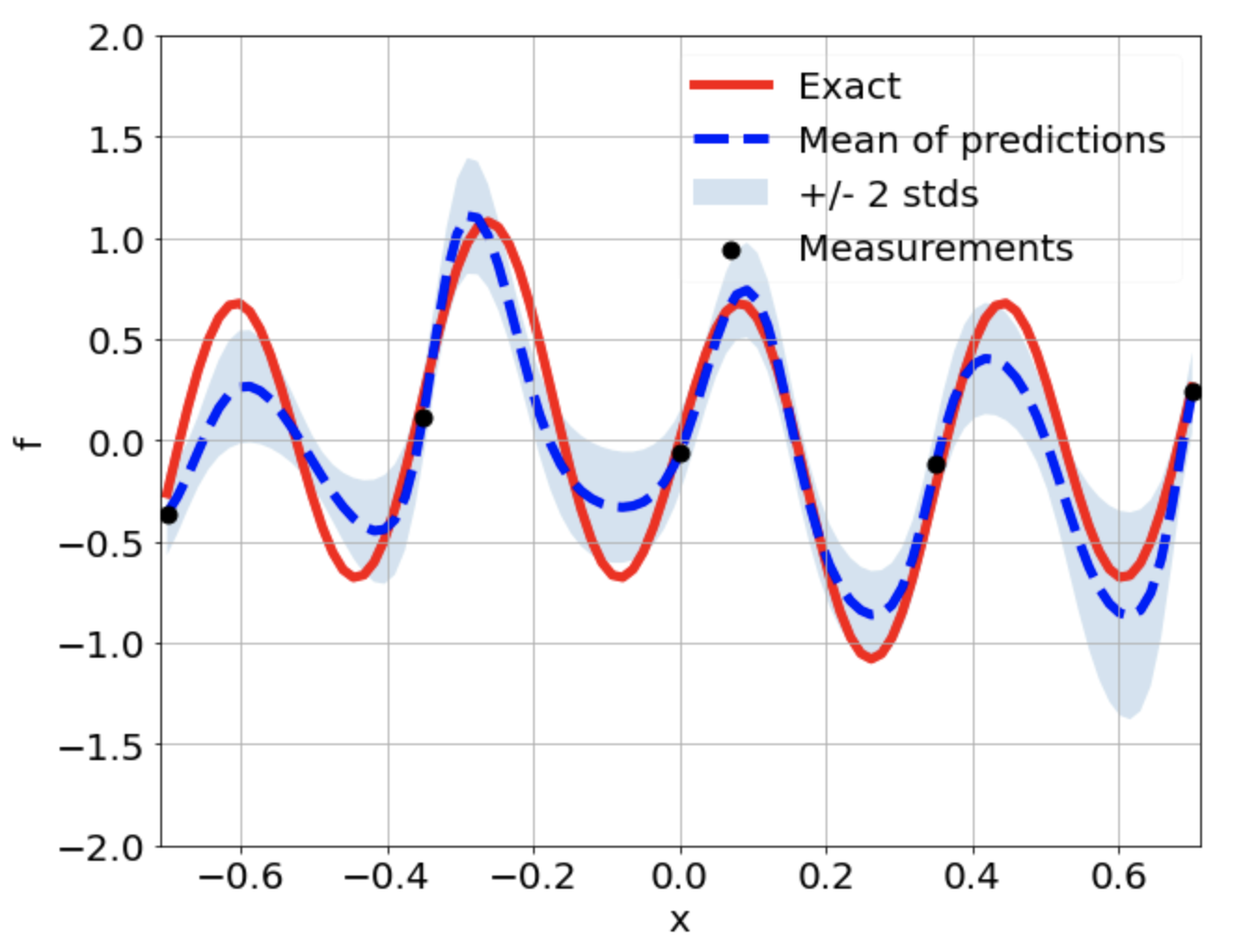} }}%
    \caption{Predictions comparison with/without using additional information.}
    \label{fig:integration improve}
\end{figure}

Next, we assume that only 5 measurements of $f$ are available and other settings remain the same. It can be seen in Figure~\ref{fig:integration improve} that the data is not sufficient to provide us with reasonable predictions of $u$ and $f$.  In practice, this is quite common that the direct measurements are expensive and sparse so that numerical simulations are usually performed to assist the predictions.  Here we assume that the mean and standard deviation of $u$ from the FEM approach at the 9 locations in Figure~\ref{fig:quantile} are also available to us and we include this additional knowledge into our training to improve our predictions.

In Figure~\ref{fig:integration improve}, it shows that the predictions of $u$ and $f$ are highly improved with using mean and standard deviation information.  Furthermore, it can be seen that the predictions can also be improved if only the means are available.  In the framework of MO-PINN, this integration of additional data can be easily implemented by modifying the loss function which makes the approach suitable for the scenarios where the sources of data are varied and complicated.

\section{Summary} \label{summary}
We proposed the MO-PINNs in this work to solve both forward and inverse PDE based problems with noisy data.  This approach allowed us to use prior knowledge of statistical distributions into the calculations under the framework of PINNs.  Our numerical results indicate that the MO-PINNs can return accurate distributions at the end of training for both forward and inverse problems. Finally, for the first time, we compared the the predicted distributions using pure deep learning framework with solutions from the traditional numerical solver (FEMs).  We also demonstrated that our approach is capable of using limited statistical knowledge of the data to make a better prediction.

In the future, we would like to further explore this approach and work on problems with multi-fidelity data which is also common in the engineering appliations.
We are also interested in using this deep learning framework with practical conventional solvers in real engineering applications which are usually very computationally expensive with the goal of make uncertain predictions faster.

\newpage
\nocite{*}
\bibliographystyle{abbrv}
\bibliography{main.bib}
\end{document}